\documentclass[final,3p,times,authoryear]{elsarticle}
\usepackage[greek,british]{babel}
\journal{International Journal of Human-Computer Studies}

\usepackage{lineno}

\usepackage{amssymb}

\usepackage{multirow}
\usepackage[caption=false,font=footnotesize]{subfig}

\graphicspath{{fig/}}
\DeclareGraphicsExtensions{.pdf,.jpeg,.png}

\usepackage{color}
\newcommand{\rvm}[1]{\textcolor{black}{#1}}

\begin{document}

%
\begin{frontmatter}

\title{A Technology-aided Multi-modal Training Approach to Assist Abdominal Palpation Training and its Assessment in Medical Education}



\author[rca,wmg]{Ali Asadipour \corref{main}}
\ead{ali.asadipour@rca.ac.uk}
\cortext[main]{Corresponding author}

\author[wmg]{Kurt Debattista}
\author[wms]{Vinod Patel}
\author[wmg]{Alan Chalmers}

\address[rca]{Computer Science Research Centre, Royal College of Art, London, United Kingdom}
\address[wmg]{Visualisation Group, WMG, University of Warwick, Coventry, United Kingdom}
\address[wms]{Warwick Medical School, University of Warwick, Coventry, United Kingdom}

%
\begin{abstract}
Computer-assisted multi-modal training is an effective way of learning complex motor skills in various applications. In particular disciplines (eg. healthcare) incompetency in performing dexterous hands-on examinations (clinical palpation) may result in misdiagnosis of symptoms, serious injuries or even death. Furthermore, a high quality clinical examination can help to exclude significant pathology, and reduce time and cost of diagnosis by eliminating the need for unnecessary medical imaging. Medical palpation is used regularly as an effective preliminary diagnosis method all around the world but years of training are required currently to achieve competency. This paper focuses on a multimodal palpation training system to teach and improve clinical examination skills in relation to the abdomen. It is our aim to shorten significantly the palpation training duration by increasing the frequency of rehearsals as well as providing essential augmented feedback on how to perform various abdominal palpation techniques which has been captured and modelled from medical experts. A comparative evaluation on usability and effectiveness of the method is presented in this study. Four professional tutors were invited to take part in the design, development and assessment stages of this study. Widely-used user centred design methods were employed to form a know-how document and an assessment criterion with help of medical professionals. Our interface was used to capture and develop a best practice model for each palpation tasks for further assessment. Twenty three first year medical students divided into a control group (n = 8), a semi-visually trained group (n = 8), and a fully visually trained group (n = 7) were invited to perform three palpation tasks (superficial, deep and liver). The medical students’ performances were assessed using both computer-based and human-based methods where a positive correlation was shown between the generated scores, $r=.62$, $p$ (one-tailed) $< .05$. The visually-trained group significantly outperformed the control group in which abstract visualisation of applied forces and their palmar locations were provided to the students during each palpation examination ($p < .05$). Moreover, a positive trend was observed between groups when visual feedback was presented, $J=132$, $z=2.62$, $r=0.55$.
\end{abstract}


\end{frontmatter}

%
\section{Introduction}
Human hands are widely used every day as exploratory and manipulative tools to obtain sensory information (e.g. shape, size, texture, roughness) and to interact with the surrounding environment. Mastering dexterous use of our hands in some motor tasks requires years of training. Significant hands-on practice is essential in particular disciplines such as medical education to master core skills (e.g. clinical palpation) and to retain them throughout a career. The highly articulated structure of the human hands provides \rvm{a range of movements designed to allow functions such as pinching, pincer grip, holding, throwing, catching, etc.} Thus, it is very hard to generalise a captured model of movements of the hand for a particular targeted skills. Moreover, various sensory receptors and their higher concentrations under the hands' tissue adds to the difficulty of studying skilful hands-on interactions.
\textbf{•}
In medical education, physical examination(s) plays a key role in diagnosing diseases from their signs/symptoms in the early stages of the patient care process. Almost all of the preliminary physical examinations involve the use of clinicians' hands and sense of touch to assist in diagnosing various conditions. Therefore it is essential that clinicians master these core skills early on in their medical education \citep{patel2011medref,dinsmore1997vrpalpation}.

A high quality clinical examination using palpation can help exclude significant pathology and reduce the need for unnecessary (expensive) imaging such as CT scanning and MRI scanning. There is at least anecdotal evidence that poorer clinical examination can result in more expensive and intrusive investigations. This paper will focus on using a novel palpation training system (PTS) to teach and improve clinical examination skills in relation to the abdomen. A patient with \rvm{an acute emergency abdominal complaint, for example perforated bowel. The latter can often be a life-threatening event.} It is also important that the clinician competently picks up any enlargement of the liver, spleen and kidneys. \rvm{The use of PTS would allow a certain degree of ability to be achieved prior to seeing many real patients in an acute or chronic management setting. PTS also allows individualised feedback and re-testing to check for improvement in practice. } The need for superficial, deep and specific organ palpation is thus essential in abdominal examination.

It is equally important that the patient is not subjected to unnecessary ``forces" when the abdomen is palpated. For example, it is unnecessary to use the heel of the hand (thenar and hypothenar eminences) as this would not contribute to any diagnostic information but may cause patient discomfort. In addition, a common error amongst novices in clinical examination is to use the tips of the finger to poke the patient rather than use the side of the hand as the diagnostic instrument.

While there is no single unique way to “correctly palpate”, a series of guidelines with suggested steps to be performed exist for correct palpation. Various resources are given to medical students by professional authorities every year (e.g. textbooks, hours of online videos, virtual reality, apps, etc.) with the major focus on how best to transfer the core skills from experts to novices, as well as revising the assessment processes to ensure that a proficient level of core skills has been attained. However, delivering these resources to a large number of students is challenging for programme directors. For instance, Warwick Medical School (WMS) has over 200 new starters per annum who are taught by over 400 senior clinicians with qualifications from 25 different countries.

Recent advances in multimodal virtual and augmented reality simulation techniques have had an impact on learning complex motor skills particularly in medical education \citep{coles2011rolehaptic}. Physical involvement in virtual simulations can deliver more immersion, minimising the risk of putting patients' health at risk. The availability of such virtual simulations is particularly valuable in early stages of medical education.

The motivation of this research is to gain a deep understanding and form a reference model for the dexterous use of hands in specific physical interactions. The goal is that training with such a derived model may help enhance the conventional motor training processes of others and allow them to be subsequently assessed via an innovative multimodal training system and not to limit them to a particular method. This is validated by a detailed study to assess the impact of our novel training approach on trainees' motor performance and learning. A mixed-mode research methodology is used to predict the usability and usefulness of the proposed technique.

In addition, User-Centred Design (UCD) research methods \rvm{such as participatory design, focus groups, evolutionary prototyping, and usability studies} were employed to ensure the new technique complies with user and domain needs. This study benefits from active collaboration with medical experts and students throughout the design and evaluation process.

This paper presents a novel technology-aided method for learning core abdominal palpation skills. A new computer-based educational interface is designed and developed to enhance learning and assessment processes in the current curriculum. A team of senior medical tutors were actively involved throughout the research and design processes to evaluate this approach and help to improve it until it satisfies their requirements. First year medical \rvm{students} were invited to use this method and to reflect on the usability and usefulness. Learning performance was significantly improved in the group of medical students who benefitted from employing this method in training and test phases ($p < .05$).



%
\section{Background and Related Works}
\label{sec:background}
In general, sensory-motor feedback is classified by research studies \citep{schmidt2005motor} into intrinsic to the individual (\textit{\textbf{Inherent Feedback}}) or could be provided as supplementary to one based on their motor behaviours (\textit{\textbf{Augmented Feedback}}). This section presents a brief overview on the second form of sensory-motor feedback which is also addressed as alternative, additional feedback.  \cite{bilodeau1966acquisition} describes augmented feedback as the most important variable in learning motor movements apart from practice. Different aspects of human hands are also presented in this section. Finally, essential parameters to capture in medical palpation training are presented with a brief review on existing data acquisition techniques.

\subsection{Augmented Feedback and Motor Learning}
Augmented Feedback (AF) is defined as the external source of information (such as a trainer or a feedback display) that helps one to understand certain aspects of motor movements and their impact of the environment within the learning process \citep{sigrist2013augmented}. AF aims to gradually minimise the variability in a closed-loop control system \citep{schmidt2005motor} via sensory feedback loop for the newly developed motor skills. Feedback helps the learner to shape complex motor skills from various components of movements by shortening the gap between actual performance and the desired skill in a self-correcting fashion within frequent rehearsals. Thus, provision of AF is widely accepted as beneficial and effective to enhance motor learning \citep{schmidt2005motor, sigrist2013augmented}.

AF is usually provided by a human expert such as a coach to the learner in form of verbal and physical demonstration of individual movements or by a technical display such as an autonomous computer-aided simulator \citep{sigrist2013augmented}. In a study led by \cite{hsiao2016using}, a Gesture Interactive Game-based Learning (GIGL) model is proposed to enhance cognitive developments and motor skills acquisition in preschool students. A ball-catching game with a vision-based gesture tracking device were used in their experiment to evaluate its impact on the students' motor control and agility compared to physically tapping a balloon in control group. Participants in the experimental group were able to adjust their body position and gross muscles based on the feedback displayed on the game screen. Significant improvement was reported in learning performance and motor skills in the experimental group.

\subsection{The Human Hands}
Human hands are always credited for their effective role on the evolution of the human species and the development of skills \citep{taylor1955anatomy}. With more than twenty seven degrees-of-freedom (DoF) in movements \citep{lin2000modeling}, human hands are considered to be the most articulated structure in one's body. Moreover, their ability to deliver high bandwidth of information through a complex network of underlying sensory organs and the size of reserved processing space in the cerebral cortex (almost equal to the total space reserved for arm, body, and legs) highlights their importance in our everyday life \citep{taylor1955anatomy}. Thus, it is essential to study their physiological and biological aspects in order to unravel the knowledge behind utilising them in complex motor skills. A brief description on mechanical and biological aspects of human hands are presented which is used as a guideline to plan for the best ergonomic approach to quantify their metrics during hands-on interactions.

\subsubsection{Mechanical Aspects} 
\begin{figure}[h]
\centering
\includegraphics[width=0.6\textwidth]{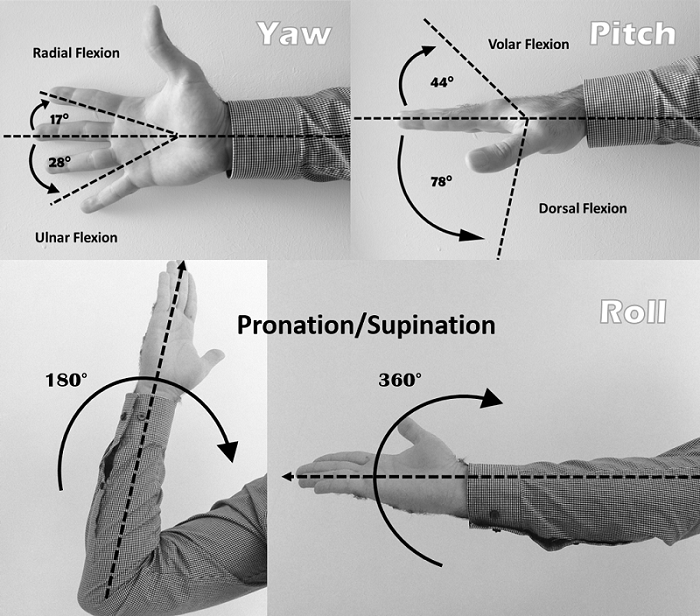}
\caption{The Human Hands' Range of Movements}
\label{fig:flexions}
\end{figure}

Despite its articulated structure, the human hands are highly constrained in range of movements. Angular displacement around wrist joints were presented by \cite{bunnell1964surgery} with total range of $122\,^{\circ}$ for \textit{\textbf{pitch}} ($78\,^{\circ}$ dorsal flexion and $44\,^{\circ}$ degrees volar flexion) and $45\,^{\circ}$ degrees for \textit{\textbf{yaw}} movements ($17\,^{\circ}$ radial flexion and $28\,^{\circ}$ degrees ulnar flexion). The rotational mobility in \textit{\textbf{roll}} action (pronation/supination movements) were also determined to be $180\,^{\circ}$ (when elbow is flexed) to approximately $360\,^{\circ}$ (with help of the shoulder when the arm is fully extended) \citep{taylor1955anatomy}. Figure \ref{fig:flexions} illustrates the angular movement constraints of the human hands.

\subsubsection{Biological Aspects}
Tactile sensations such as vibration, touch and pressure are perceivable by the underlying network of sensory receptors called mechanoreceptors. 
\cite{purves2001mechanoreceptors} highlight the mechanoceptors locations in different layers of the human hands' tissue. Fingertips have the highest discriminative capacity with two-point threshold (minimum required distance to discriminate two stimuli) of $2\ mm$ that increases to $10\ mm$ on the palm with higher sensitivity threshold of $2\ gm/mm^2$ as compared to the forearm $33\ gm/mm^2$ \citep{taylor1955anatomy}. Characteristics of other types of tactile receptors such as \textit{Thermoreceptors} (temperature) and \textit{Nociceptors} (pain) are out of scope of this study and so have not been reviewed.

\subsection{Palpation Metrics and Quantification Techniques}
In order to study abdominal palpation examinations it is crucial to first identify which key parameters need to be measured. Since the global location of the patient's abdomen remains stationary during the examination, it is only necessary to identify metrics that are related to the human hands. Therefore, three important parameters are proposed by both medical and non-medical researchers \citep{bendtsen1995pressure,williams2012hand,prisacariu2011robust,wang2009real,oikonomidis2012tracking} as outlined below:

\begin{itemize}
\item \textbf{\textit{Position:}} palpating hand's position on the patient's abdomen
\item \textbf{\textit{Orientation:}} rotational displacements of the palpating hand
\item \textbf{\textit{Pressure:}} magnitude of the exerted forces by different regions of the palpating hand on patient's abdominal surface.
\end{itemize}

The first two parameters are usually referred to as \textit{hand tracking metrics}. The third parameter was explored by ergonomic studies on the human haptic system to unravel the wisdom of dexterous application of force within physical interactions.

\subsubsection{Hand Tracking}
Tracking the human hand is a challenge due to its highly articulated structure \citep{taylor1955anatomy,oikonomidis2012tracking}. The measurement process in hand tracking comprises global tracking which indicates the position of the whole hand as a structure, and posture tracking which indicates the formation of hand kinematics (bones and joints) in 3D space. In general, two types of tracking methods are possible; vision-based and wearable-based \citep{lin2000modeling}. The former is mostly covered by visual motion tracking techniques such as camera-based tracking solutions (eg. Microsoft Kinect, Leap Motion, Sony EyeToy etc.) in a marker-based (using a set of sensors or markers) or marker-less fashion with help of image processing algorithms \citep{frati2011using,tang2011recognizing,prisacariu2011robust}. 

\rvm{A novel real-time approach is proposed by \cite{sridhar2016real} for simultaneous hand-object tracking using a 3D Gaussian mixture alignment strategy. A three step pipeline is designed and developed with a two-layer part classification strategy. The authors captured a new benchmark dataset for both fingertip positions and object pose to handle occlusion difficulties. However, the proposed technique will not perform efficiently where the hand is occluded for a long period, and for complex object shapes and fast interactions.}

The latter employs a set of sensors (gyroscope, accelerometer, flex, magnetic etc.) which are attached to either a glove or fixed by using straps to the hand in order to estimate the hand positions/orientations in 3D space without need of any visual based tracking \citep{dipietro2008glovesurvey}.

\subsubsection{Pressure Mapping}
Various types of force sensors/gauges (mechanical, resistive-based, capacitive-based, magnetic-based and etc.) were used in different studies on the human ergonomics to measure the variations in applied forces in physical interactions \citep{bendtsen1995pressure,futarmal2011palpometer,spyridonis2010back,van2009neck}. In a study led by \cite{bendtsen1995pressure}, a Pressure-Controlled Palpation (PCP) technique was introduced to monitor the applied pressures by a participant's right index fingertip. Variations in applied pressures by different individuals in absence of a standardised model was suggested as a serious problem in learning muscle palpation examinations. Thus, a novel quantification tool (\textit{Palpometer}) was developed and proposed to study these variations. Significant variation was reported between participants in palpation of temporal muscle and mastoid (different persons - different tasks) when no visual feedback was provided and these variations were paramount among inexperienced subjects. Women participants had the lowest force readings as compared to men.

In another study by \cite{futarmal2011palpometer}, a low-cost mechanical palpometer were used to facilitate the assessment of deep pain sensitivity in myofacial palpation. Clinicians were able to feel the tapering end of the mechanical rod when it was fully pressed. Three springs were used ($0.5, 1.0,\ $and$\ 2.0$ Kg) to provide resistance against mechanical displacements of the rod inside the cylinder to provide different force levels (springs were calibrated by a digital force meter). \cite{futarmal2011palpometer} have reported a low test-retest variability with more accurate control on force when mechanical palpometers were provided.

The results from previous studies presented in this section were used in our study to enhance the robustness and reliability of the measurement technique that is proposed in section \ref{sec:interface}.


%
\label{sec:ergonomics}

\section{Methodology}
\begin{figure}[htbp]
\centering
\includegraphics[width=0.6\textwidth]{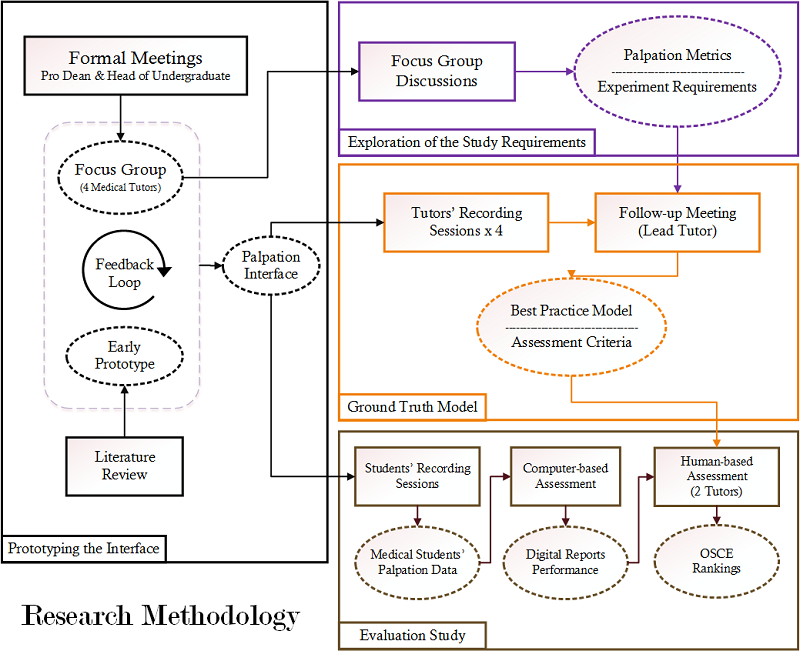}
\caption{Research Methodology - procedures are presented by solid rectangles and predicted outcomes are illustrated by dashed ellipses}
\label{fig:reme}
\end{figure}

Elicitation of user requirements (or extracting human factors) is a compulsory step to predict the aspects of usability and usefulness of a technique in human-centred research studies \citep{martin2012user}. Thus, qualitative and quantitative data have been collected in different stages of this research via a series of User Centred Design (UCD) methods. Also, to elaborate the user and domain needs, a team of medical tutors and their first year students were actively involved in this study. Figure \ref{fig:reme} shows the step-by-step design of this research methodology.

The approach proposed in this paper has been developed in consultation with experts with specific emphasis in improving clinical competency in palpation \rvm{(use of the correct part of the hand to aid the quality of palpation and not cause patient distress)} in order that no significant intra-abdominal pathology is missed, whilst still preserving patient dignity and comfort. Three key steps are defined by the senior clinical tutors to ensure that each student participant was able to examine the abdomen with the correct amount of pressure for competent superficial and deep palpation. These formed the basis of our methodology: 
\begin{itemize}
\item Full coverage of the abdominal surface,
\item Differentiation between superficial and deep palpation, and
\item Prevention of leaning on abdominal surface with thenar and hypothenar eminences 
\end{itemize}

Different stages of this research methodology are briefly discussed in this section. More detail about each stage is presented within its representative section.

\subsection{Prototyping the Educational Interface}
Design of an early prototype for palpation training was planned due to a lack of available off-the-shelf measurement interfaces with the flexibility required for this study. Moreover, additional factors such as budget restrictions and time commitments were influential in making this decision. \textit{\textbf{Cognitive Walkthrough}} was used to evaluate the early prototype, when an end-user is not available with the help of proxies (e.g. fellow colleagues) to ensure it complies with the domain principles which were extracted via literature reviews and process observations.

The early version of the prototype was made to demonstrate the concept of a technology-aided training interface. It was evaluated by a team of medical tutors (Focus Group) to identify a list of design refinements before the data collection stages. 

\subsection{Exploration of Study Requirements}
\label{subsec:requirements}
In a follow up communication with Pro Dean and head of the undergraduate studies (Bachelor of Medicine, Bachelor of Surgery) a team of medical experts who teach physical examinations within a current curriculum was identified.

\textit{\textbf{A Focus Group (FG)}} method was used with a seat-back strategy (collecting data by listening to the group discussions) to extract experiences via discussions between medical tutors and to identify current challenges in the abdominal palpation examinations educational process. User and task requirements were identified and transcribed in the form of meeting protocols (or minutes) from verbal discussions and written answers in the tutor’s questionnaire as part of the self-reporting technique.

\subsection{Creation of a Proof of Concept Model}
The knowledge of the dexterous use of the human hands within abdominal palpation examinations is usually transferred from masters to novices via lifelong rehearsals of clinical skills \citep{patel2011medref,macleod2009macleod}. Thus, it is essential to model the medical tutor’s palpation skills via a novel quantification technique in order to establish a ground-truth reference. The existence of such a reference model is needed in the evaluation stage of this research to ensure all students are going to be assessed in exactly the same manner. Palpation metrics (position, orientation, pressure) were collected via an innovative measurement interface (see section \ref{sec:interface}) from the medical tutors' sessions. In this paper we focus on applied pressures at different regions of the human hands which were analysed and employed in the evaluation study. Section \ref{sec:ground-truth} presents detailed overview on proof of concept stage and the potential findings from the tutors' performances.

\subsection{Evaluation Study}
\textit{\textbf{A Usability Test}} method was used in the final stage of this study to evaluate if the proposed method can fulfil its potentials in learning complex motor skills. The impact of real-time multimodal augmented feedback of applied forces on the medical students' competency was assessed by this experiment. Consequently, a digital competency report is provided to evaluate its benefits in the medical tutor’s assessment process. Therefore, a pilot study was scheduled in the last FG session to employ this technology-aided training and assessment technique in one of the students’ formal training sessions to evaluate its usability and usefulness in a real-world practice. Quantitative measurements were collected from students' performances to be compared with the best practice model for each abdominal palpation task during the assessment process. In addition, quantitative data were collected from students who had had a chance to learn abdominal palpation skills with the new technique via usability feedback forms, section \ref{sec:evaluation}.

%

\section{Study Requirement}
Domain requirements were identified in the Focus Group discussions via open-ended questions and carefully noted from the discussion. These recommendations were employed throughout the interface prototyping, experiment design and data collection phases.

\subsection{User Requirements}
The following user requirements were identified as key concerns to accept technological interventions in medical education:

\begin{itemize}

\item \textit{\textbf{Complementary Method}} - the role of medical tutors in conventional medical training is undeniable and could not be replaced. Hence, new technology-aided interventions should be used in conjunction with the current practice in the medical curriculum.

\item \textit{\textbf{Real-time Feedback and Instant Assessment}} - instant feedback on competency is highly demanded during the learning phase. Also, an instant report on students' overall performance could enhance the existing assessment process.

\item \textit{\textbf{Unintrusive Design}} - the interface must provide a similar experience to the real-world experience without any interference in the palpation process.

\item \textit{\textbf{Health and Safety}} - when wearable technologies are in direct contact with human skin, constant use by different medical users on several patients may result in skin infections. Also, an electrostatic shock could occur and it could put the patient's safety at risk. In addition, exhaustion caused by the measurement interface in quantification studies could directly affect the user's competency and experience.

\item \textit{\textbf{Time and cost}} - these are often identified as the two paramount challenges in the adaptation of a new technology-aided intervention in current medical education.

\end{itemize}

\subsection{Task Requirements}
Three abdominal palpation examination tasks were selected to investigate two important learning objectives; to highlight the transition between applied forces (superficial vs. deep palpation) and to illustrate the correct formation of the palpating hand by localisation of the applied forces (locate specific organ e.g. liver).

\begin{figure*}[t]
\centering
\includegraphics[width=0.9\textwidth]{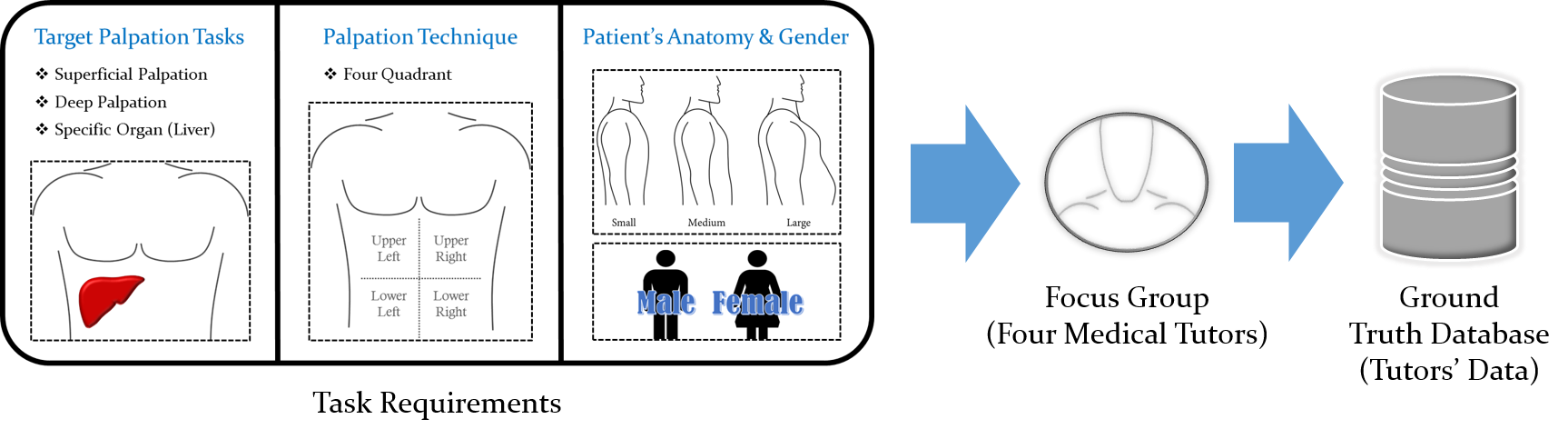}
\caption{Task requirements}
\label{fig:task-reqs}
\end{figure*}

Palpation of four quadrants with right hand was chosen by the medical tutors to minimise variations caused by employment of different techniques for superficial and deep palpation tasks. Actor patients were invited from two genders with three different anatomical variations to ensure the proof of concept phase outcomes were reliable. Figure \ref{fig:task-reqs} shows the task requirements that were identified by the medical tutors in group interviews.

%

\section{Palpation Training System (PTS)}
\label{sec:interface}
The initial wearable prototype which is used in this study was the lightest (70 grams), thinnest (reading pulses on fingertips), and cheapest version of the data gloves  with 12 sensors (three on the radial border of the hand) and long range high speed wireless communication over Bluetooth. An additional essential requirement in clinical practice is that the glove has to be able to withstand alcohol sanitising gel for purposes of high standards of hand hygiene \citep{safety2009guidelines}. Detailed information about characteristics, fidelity and calibration processes of the sensors are provided in \citep{asadipour2016visuohaptic}.

\subsection{Current state of the art}
High prices per unit (eg. extra charges for software and more sensors), long calibration processes, lack of flexibility in design, poor coverage, and bulkiness are some of the disadvantages of existing commercial quantification tools. Moreover, the majority of the commercial products are generic without active involvement of the end-users. To overcome these limitations, an innovative low-cost measurement interface was designed and developed with the guidance of medical experts to satisfy all of the study requirements, as discussed in the previous section.

\subsection{Design}
\begin{figure}[htbp]
\centering
\includegraphics[width=0.6\textwidth]{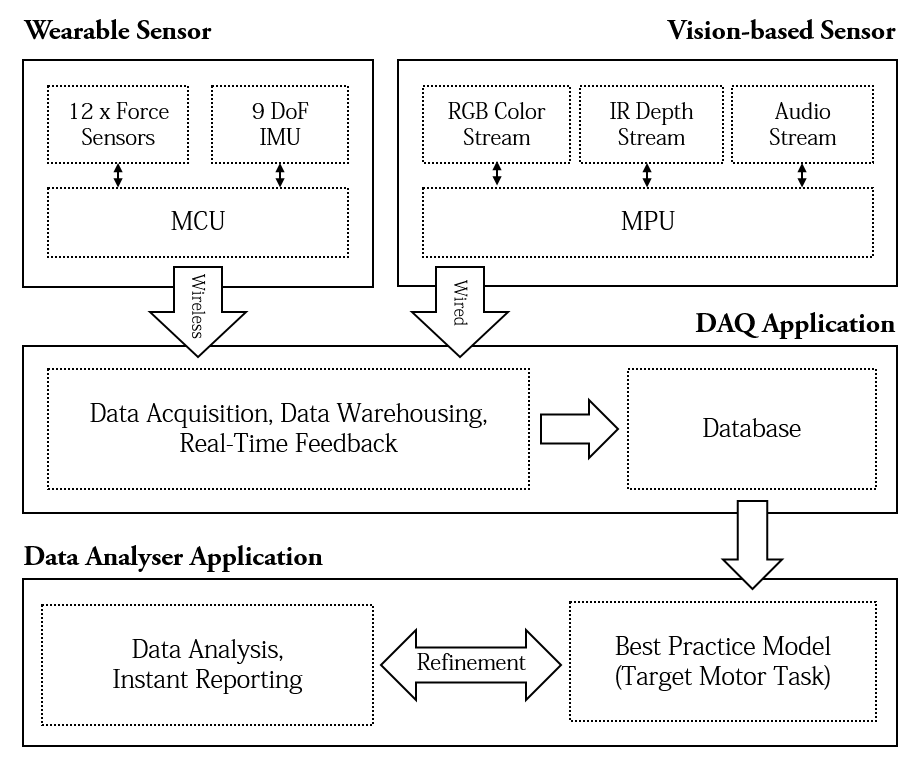}
\caption{PTS system block diagram}
\label{fig:pipeline}
\end{figure}

The PTS interface layout and its outputs in different stages of this research are highlighted in Figure \ref{fig:pipeline}.

\subsubsection{Multidisciplinary Design Team}
An early version of the measurement interface was evaluated by each member during the first FG session and design refinements were summarised by the interaction between the medical tutors. One tutor who currently teaches abdominal palpation skills in WMS was invited in a series of follow up meetings throughout the iterative design process, to advise on detection of potential patterns and knowledge from the captured data from the medical tutors, data visualisation and further analysis. The medical tutors' feedback in the design stage was crucial to create assessment criteria and to generate competency reports prior to the evaluation study.

\subsubsection{The Interface Framework}
\paragraph*{\textit{\textbf{Pressure:}}}
a common technological challenge in force quantification studies to measure applied forces during hands-on interactions is the installation of the transducers (e.g. force sensors) on such an articulated structure (the human hands) without decreasing dexterity \citep{jensen1991conductive}. A series of contact points on the human hand for particular interactions are used as a reference to mount force-sensing transducers with smaller spatial resolution to enhance the flexibility of the hands.

\begin{figure}[htbp]
\centering
\includegraphics[width=0.4\textwidth]{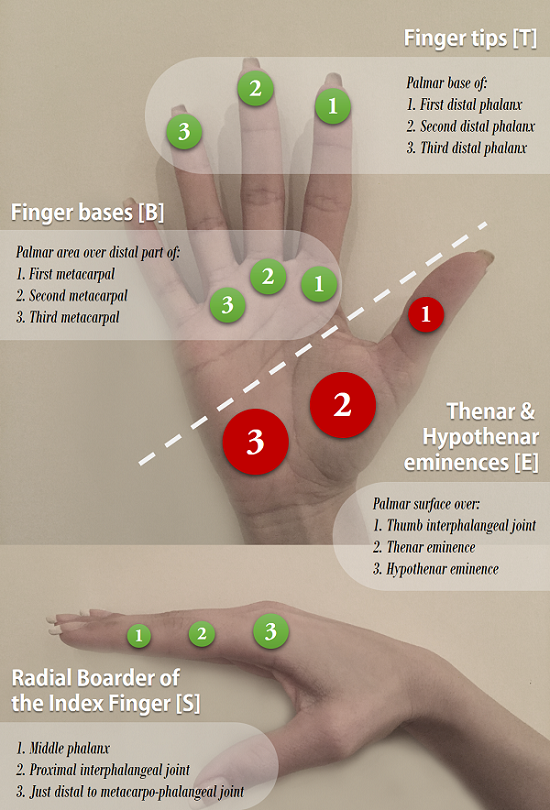}
\caption{Contact points on the palpating hand's surface}
\label{fig:hand-sensor}
\end{figure}

Twelve contact points were identified on the right hand's palmar surface and its radial border of the index finger, based on advice given by the medical tutors in the user requirements elicitation phase. These points were identified during live demonstrations of target abdominal palpation tasks (superficial, deep, and liver edge) on a dummy mannequin with the initial prototype of PTS. Figure \ref{fig:hand-sensor} shows the contact points' positions on the right hand as the palpating hand, for all three tasks.

\paragraph*{\textit{\textbf{Orientation:}}}
another important metric to be measured during the abdominal palpation examination process is the rotational movements of the palpating hand in space (see figure \ref{fig:flexions}). In order to account for all useful orientations, a wearable sensory quantification tool in the form of a glove was proposed as part of the PTS interface to measure applied pressures as well as the rotational movements of the human hand during palpation process.

\begin{figure}[htbp]
\centering
\includegraphics[width=0.6\textwidth]{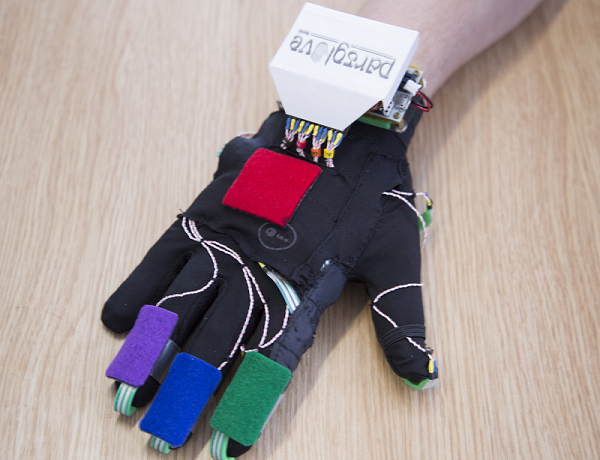}
\caption{ParsGlove is the wearable sensory input with colour markers to capture applied forces, rotational movements and spatial position of the human hand}
\label{fig:pars-glove}
\end{figure}

The force and orientation metrics were digitally sampled by two on-board micro-controllers and the raw data \rvm{is} transmitted from a wearable technology (\textit{ParsGlove\texttrademark}) to the base computer via a wireless solution to provide freehand interactions in a non-intrusive fashion. Figure \ref{fig:pars-glove} shows \textit{ParsGlove\texttrademark}, the prototype based on the feedback from the medical tutors.

\paragraph*{\textit{\textbf{Position:}}}
the position tracking techniques are highly influenced by the resolution of the target tracking area on the human anatomy. Vision-based tracking techniques were surveyed to identify the best approach which complies with the user requirements in this research. A marker-based tracking method, comprising four colour markers, is used to capture global (palmar) and local (fingers) movements of the human hand since a wearable sensory input is already introduced to capture other metrics. Colour Markers were attached to the ParsGlove on areas which were defined by the medical tutors. A Kinect sensor was mounted above the patient's body and was focused on the abdominal region to detect the markers and return three dimensional values to represent position of each marker in space. A point cloud is also captured in each frame for further work on the anatomical variations to form an average body model and locate organs.

\subsubsection{PTS Data Acquisition Application (PTS-DAQ)}
\begin{figure}[htbp]
    \centering
    \subfloat[Monitoring Panel]{\label{fig:a}
    \includegraphics[width=0.6\textwidth]{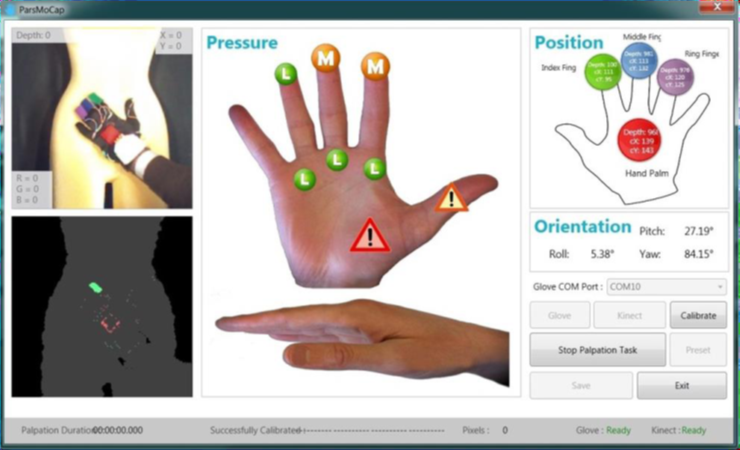}}
    \subfloat[Feedback panel]{\label{fig:b}
    \includegraphics[width=0.305\textwidth]{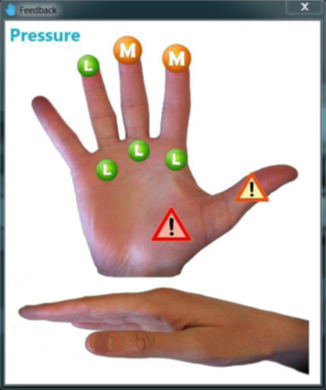}}
    \caption{PTS-DAQ; PTS data acquisition user interface}
    \label{fig:pts-daq}
\end{figure}

The data acquisition user interface is comprised of two panels. The main panel (\textit{Monitoring Panel}) monitors the data collection process and an auxiliary panel (\textit{Feedback Panel}) provides visual feedback on applied pressures on the patient's abdomen by the medical user (locations and magnitudes). Different force magnitudes are represented by three colour-coded symbols ranging from small (green), medium (amber) and high (red) based on the information extracted from the tutors' best practice model. The captured information could be stored on a local drive (Excel spreadsheet) or on an online database. Figure \ref{fig:pts-daq} shows the PTS data acquisition user interface.

\subsubsection{PTS Data Analysis Application (Examogram)}
The data analysis user interface (hereafter known as Examogram) was designed, based on the requirements that were explored in series of meetings with the lead medical tutors. These included the ability to perform preliminary analysis such as frequency of presses per task, the highest recorded pressure per press, and the duration of each press on the tutors' datasets to create a proof of concept model. Examogram was also used to perform similar statistical analysis and generate digital competency reports on the medical students' datasets. These outcomes were used to compare the students' competency against the proposed best practice model as part of the competency assessment process.

\begin{figure}[htbp]
\centering
\includegraphics[width=0.8\textwidth]{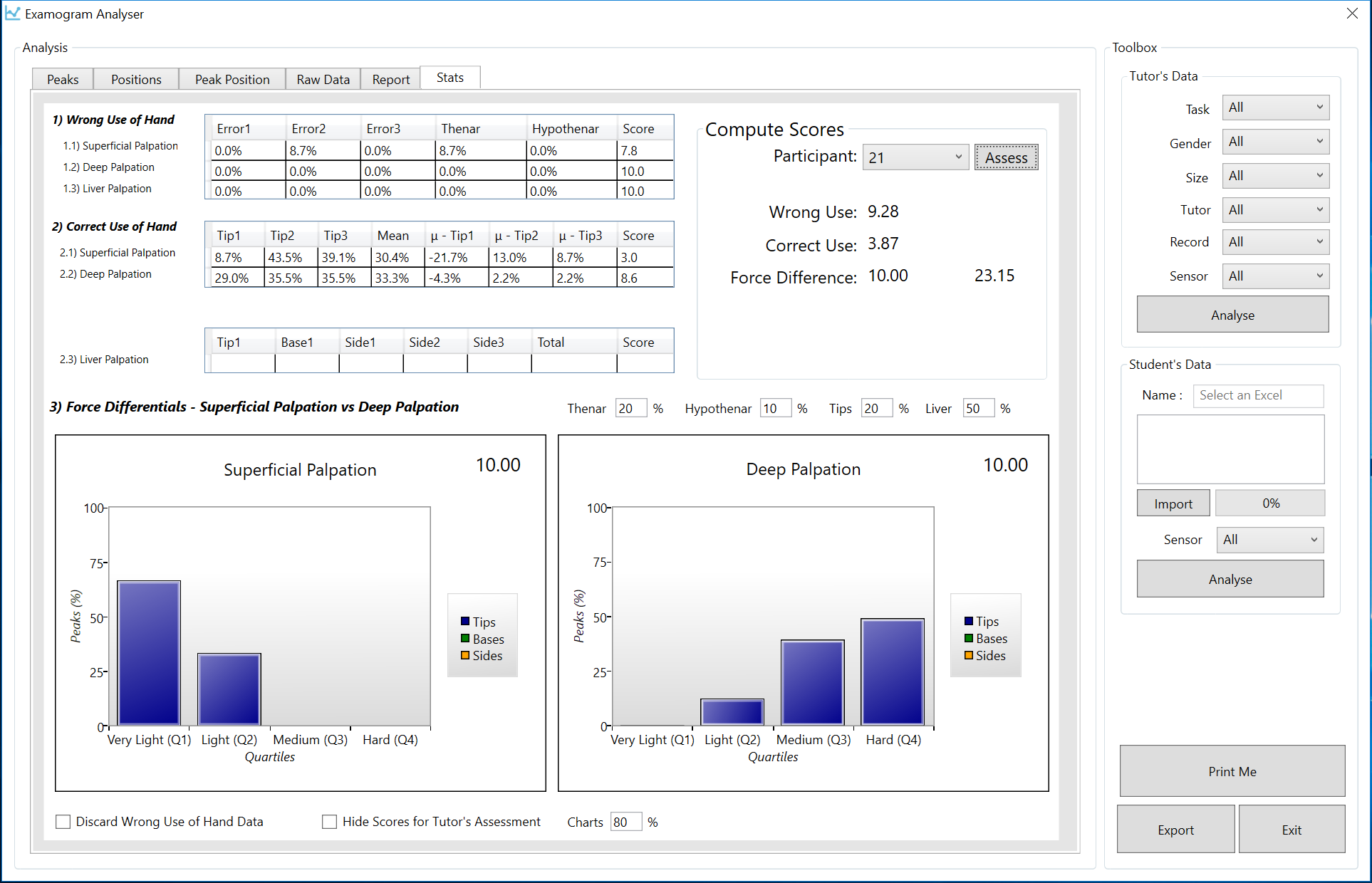}
\caption{Examogram; PTS data analyses user interface}
\label{fig:examogram}
\end{figure}


%
\section{Proof of Concept}
\label{sec:ground-truth}
This section describes the open-ended exploration and data collection method used to form an understanding of palpation activities based on the capture of medical palpation experts, from different backgrounds, across a number of palpation tasks.

\subsection{Design}
As noted in section \ref{sec:ergonomics} four medical tutors experienced in teaching abdominal palpation examination skills were invited to take part in this research. A theoretical sampling method is used to develop a well-rounded theory with the help of experienced participants using on-going interpretation of data to identify target information to collect and to generate key concepts via observations and a series of open-ended questions. The tutors, each with different medical specialities, were invited to explore different perspectives on similar subjects. Three abdominal palpation tasks (superficial, deep, and liver) were chosen by medical tutors to investigate various aspects of abdominal palpation examination via hands-on practice. Superficial and deep palpations were chosen to study the transition of force application from soft and moderate presses to much harder presses in the deep palpation case. Liver palpation was selected to evaluate correct formation of the hand and application of force only by specific parts of the hand (radial border, index fingertip and index finger base).

The interface capabilities were demonstrated to all tutors prior to their data collection session. A list of minor modifications were suggested by the medical tutors in the initial research and design meeting. The measurement interface and its software applications were reviewed based on this list to apply necessary changes. Potential risks in measurement outcomes were discussed in a series of meetings before the data collection session in this stage. Anatomical variations between actor patients and employment of different techniques were taken into account. Three different body types in two genders were identified and invited to take part in this study. Also, an averaging method with help of the medical tutors was employed to resolve  the variations.

\subsection{Materials}
The PTS data acquisition user interface is used to capture three important parameters from the medical tutors. A TFT display was positioned in front of the investigator to let him monitor the data collection process but no visual feedback was provided to the tutors or to the actor patients.

The Examogram user interface was used to analyse the medical tutors' raw data. The analysis results were used to develop a ground-truth model for each palpation task, subsequently used in evaluation experiment in the next section.

Examination and data capture took place on a mattress provided in the lab. The mattress was covered with a disposable couch roll for each participant and a medically approved hand sanitiser gel was provided as well as ultra thin powder coated polyvinyl gloves to be used inside the wearable interface to meet hygiene and safety requirements.

\subsection{Participants}
WMS medical professionals were invited to take part in this study. The team comprised three male and one female tutor with different medical backgrounds and one Gastroenterologist. All had extensive experience in acute and non-acute abdominal examinations in various setting.

Actor patients were recruited in this study as substitutes for real patients. The term actor patient refers to a normal human in average medical condition who has no known disease or abnormality at the time.
Five actor patients took part in this experiment, with ages from 24 to 31 years old.  Since this research aims to propose a reliable model for abdominal palpation examination, actor patients were selected with respect to anatomical variations and genders. Thus, body Mass Index (BMI) was used primarily to find potential participants for the three groups: Small, Medium, and Large (underweight, healthy weight, and overweight). It was difficult to find a female participant in the Large category as no one within that category volunteered and it was deemed unethical to ask individuals directly.

Other medical indexes such as Body Adiposity Index (BAI) and Waist to Hip Ratio (WHR) were also employed to calculate excess abdominal fat concentration. Table \ref{table:body-mass} shows three commonly used indexes to assess healthiness for participants which were employed during the experiment.

\begin{table}
\centering
\def\arraystretch{1.25}
\caption{Computed health metrics for the actor patients}
\label{table:body-mass}
\begin{tabular}{|l|l|l|l|l|}
\hline
Gender                  & Size   & BMI  & BAI    & WHR  \\ \hline
\multirow{3}{*}{Male}   & Small  & 20.3 & 16.9\% & 0.91 \\ \cline{2-5}
                        & Medium & 23.6 & 21.0\% & 0.97 \\ \cline{2-5}
                        & Large  & 29.5 & 25.6\% & 1.01 \\ \hline
\multirow{2}{*}{Female} & Small  & 16.2 & 22.7\% & 0.85 \\ \cline{2-5}
                        & Medium & 18.6 & 23.3\% & 0.86 \\ \hline
\end{tabular}
\end{table}

The above metrics (BMI, BAI and WHR) can provide rough estimates of the participants' body type \citep{nhs2015bmi}. The healthy weight region in BMI ranges between 18.5 to 25. Body Adiposity Index (BAI) was also used to determine the percentage of body fat from size of hip circumference and height (ShapeSense, 2015). The healthy region for this indicator is deﬁned between 21\% to 33\% for a female and 8\% to 21\% for a male (age 20 - 39). Finally, Waist to Hip Ratio (WHR) is also used by some doctors to determine if a person is carrying too much weight around his/her abdomen. A ratio above 0.85 for females and 1.0 for males indicates the risk of abdominal obesity in future.

The final decision for assigning the participants in the three anatomical categories based on their abdominal obesity was made by the medical tutors. 
Just prior to the start of the data collection, the male participant in the medium body category withdrew. 

The reason to choose participants from different genders and various body types is to develop a technique to build an average patient anatomy as well as providing options to be selected by tutees in the future. Another observation from clinical teaching is that students are apprehensive when examining obese patients because they feel that they have to “dig deeper” to overcome the subcutaneous fat to be able to palpate deeper structures such as liver and kidneys. This fact is not investigated further in this paper.

\subsection{Procedures}
Four different sessions were scheduled according to the tutors' and participants' availability. The tutors were formally invited to take part in this experiment via email invitations with enclosed ethics documentation. The actor patients also received an invitation email with their specific version of the ethics documentation.

All participants were introduced to the tutor in an induction meeting prior to the data collection. Participants were asked to attend in sequential order. It was estimated that the tutors would take twenty minutes for each participant based on five minutes per task with two and a half minutes rest intervals between tests. Also, five minutes preparation time was permitted between participants. In the preparation intervals tutors had five minutes to put on a polyvinyl glove, the ParsGlove, and to clean the ParsGlove surface with sanitiser gel. Meanwhile, the investigator was responsible for preparing the interface and checking the captured data. Tutors were asked to perform the target palpation tasks (superficial, deep and liver) twice to guarantee sampling reliability.

The tutor and actor patient were surrounded by a curtain during examination and the software application was focused only on the actor patient's abdomen (area between participant's ribs and hip). During palpation tutors were positioned in the right hand side of the patient using their right hand to palpate. Target tasks shared a common entrance point: from the right hand side of the patient's abdomen but the liver palpation had a different movement pattern and hand formation. Palpation was started with a brief introduction from tutors and a short description of the palpation routine to the actor patient before each task. Data collection was initiated and terminated by the tutor's verbal signal for each task.

\subsection{Assessment Framework}

\begin{figure}
\centering
\includegraphics[width=0.8\textwidth]{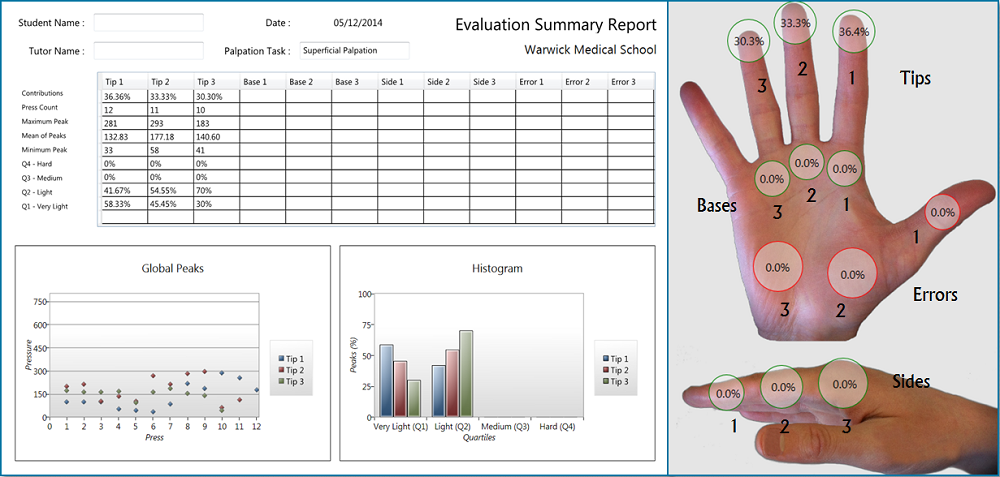}
\caption[Preliminary version of reports to design the assessment criteria]{An example of the medical students' Preliminary reports to design the assessment criteria - superficial palpation}
\label{fig:1st-rep}
\end{figure}

The captured data was analysed and was used to develop an assessment framework for subsequent use in evaluation. The medical tutors' palpation data were analysed via the \textit{Examogram} user interface and electronic reports were generated to highlight force properties such as magnitude and distribution across the hand surface and around its edge. Figure \ref{fig:1st-rep} shows an electronic report for a superficial palpation.

The raw pressure values were initially analysed in an arbitrary unit format based on 10 bits of data ($0-1023$). The applied force upper-bound threshold ($600$ arb. unit) was chosen from the tutors' data in the deep palpation examination task and this region was divided into four quartets to represent very light ($Q_1$), light ($Q_2$), medium ($Q_3$) and hard ($Q_4$) presses. Figure \ref{fig:quartiles} shows the medical tutors' raw data on a histogram in superficial and deep palpation tasks.

Force distributions were computed and visualised on a palpating hand figure. Each sensor's contribution ($C_{sensor}$) in the palpation examination process is computed as a function of the total captured press counts from that sensor ($PC_{sensor}$) and the total captured press counts ($PC_{total}$) from all twelve sensors as $ C_{sensor} = (PC_{sensor} / PC_{total}) \times 100 $.

Press frequencies and maximum force applied per press are calculated per sensor on the first version of the electronic reports. This quantity is used in the second set of reports to score students' performance in each criterion both in computer-based and human-based assessment methods. Reports were anonymised and presented to a lead medical tutor in a follow up meeting to interpret and design an assessment framework for computer-based assessment. Three criteria were defined by the medical tutors to mark students' performance. Each criteria is allocated ten points for a maximum score of thirty.

\paragraph*{\textbf{Wrong Use of Hand:}} Thenar and Hypothenar eminences (see figure \ref{fig:hand-sensor}) on the palmar surface of the palpating hand are not supposed to be used on the patient's abdominal surface during the palpation examination process, to avoid patient discomfort and possibly injury especially when they are experiencing pain. Hence, hereafter we call the sensors that detect these actions, non-permitted or error sensors ($E_i$). This criterion should be met in all three tasks.

\begin{itemize}
  \item Thenar eminence ($E_1 + E_2$) sensor's contribution (see Eq. \ref{equ:thenar}):
  \begin{equation}
    C_{(E_1)} + C_{(E_2)} \leq 20\%
    \label{equ:thenar}
  \end{equation}
  \item Hypothenar eminence ($E_3$) sensors' contributions (see Eq. \ref{equ:hypothenar}):
  \begin{equation}
    C_{(E_3)} \leq 10\%
    \label{equ:hypothenar}
  \end{equation}
\end{itemize}

\paragraph*{\textbf{Correct Use of Hand}} a good use of the hand refers to an equal balance between applied forces on the three fingertip sensors ($T_1,T_2,T_3$) due to the hand's posture in superficial and deep abdominal palpation. However, in palpation of the liver edge, it is important to employ only five sensors out of the nine permitted ones. These sensors are located on the radial border of the index finger (three side sensors), the index finger-tip and the index finger-base ($S_1, S_2, S_3, T_1, B_1$).

\begin{enumerate}
  \item Superficial and Deep palpation: the variation ($\delta_{(T_i)}$) of each fingertip sensor's contribution ($C_{(Ti)}$) from the fingertip sensors' mean ($\mu$) are computed ($\delta_{(T_i)} = C_{(Ti)} - \mu$). Ten points were given when the differences are within an acceptable threshold (see Eq. \ref{equ:delta}).
  \begin{equation}
    \delta_{(T_i)} \pm 20\%
    \label{equ:delta}
  \end{equation}

  \item Palpation of liver edge: Ten points were awarded if over half of the sensors' contribution in the examination process are focused on the radial border, finger-tip and finger-base of the index finger (see Eq. \ref{equ:liver}).
  \begin{equation}
    C_{(S_1)} + C_{(S_2)} + C_{(S_3)} + C_{(T_1)} + C_{(B_1)} \geq 50\%
    \label{equ:liver}
  \end{equation}
\end{enumerate}

\paragraph*{\textbf{Force Magnitude Transition}} four different force magnitudes were defined based on the tutor's best-practice model: very light, light, medium, and hard press ($Q_{1}$ to $Q_{4}$). In superficial palpation the main objective is to perform a gentle examination, covering all regions on the patient's abdomen. While performing deep palpation the medical students must apply harder presses in order to locate organs or to identify abnormalities. Therefore, in the superficial palpation examination the maximum peak of each press must be in the very light or light quartets whereas in deep palpation examination they need to be in medium to hard quartets.

\subsection{Findings and Discussion}
The analysed results were used by the tutors to design a framework for further palpation performance assessments. Variations caused by the tutors and techniques were  investigated and interesting patterns identified in the medical tutors' dataset. Indeed, quantifying the force needed to appreciate a pathology via a big sample of medical professionals could help determine the required thresholds to detect that pathology on a wide variety of anatomies. 

\subsubsection{Variations across Medical Professions}
\begin{figure}[htbp]
\centering
\subfloat[Superficial Palpation]{\label{fig:a}\includegraphics[width=0.5\textwidth]{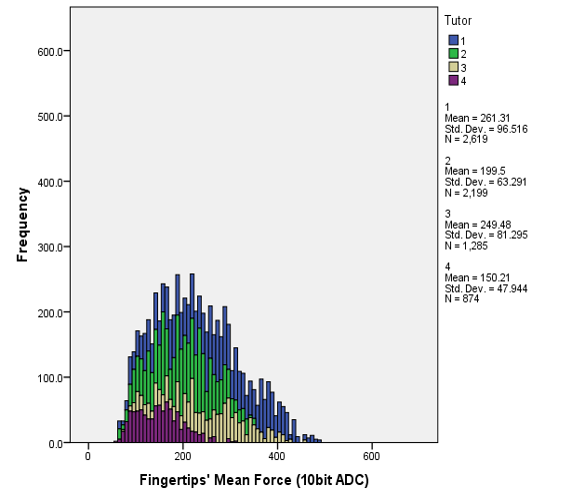}}
\subfloat[Deep Palpation]{\label{fig:b}\includegraphics[width=0.5\textwidth]{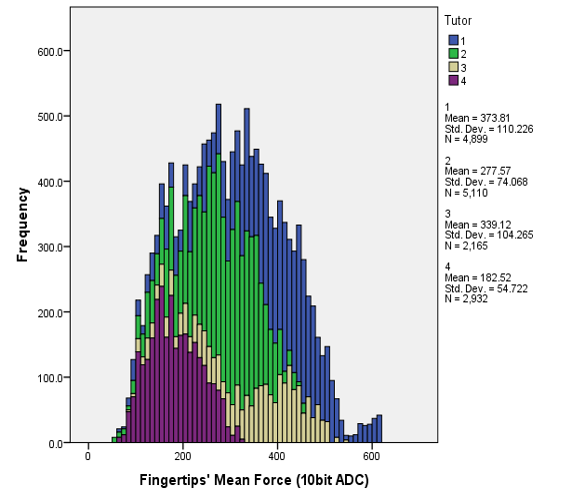}}
\caption{Fingertips' mean force when all three were engaged; rightward shift towards higher force exertions by the medical tutors in deep palpation (b) compared to the superficial palpation (a).}

\label{fig:quartiles}
\end{figure}

\begin{figure}[htbp]
\centering
\subfloat[Lead Medical Tutor]{\includegraphics[width=0.5\textwidth]{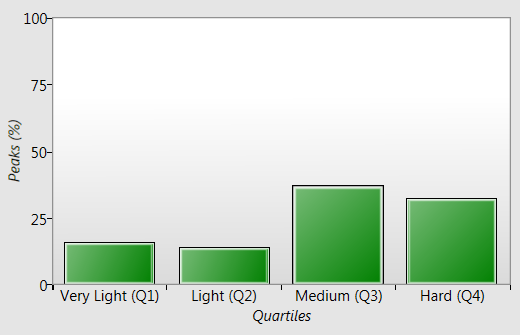}}
\subfloat[Gastroenterologist]{\includegraphics[width=0.5\textwidth]{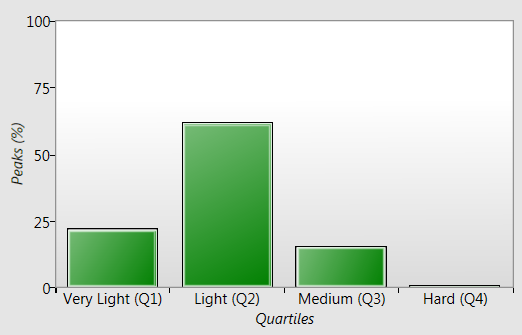}}
\caption{Variation across medical professionals in deep palpation task}
\label{fig:journey-master}
\end{figure}

Since the variation between tutors and their techniques is a well-known challenge in the medical domain, the best practice model is based on averaging the tutors' overall performance to cope with this problem. In addition, variation between medical professionals could have an effect on their force application magnitude. According to the lead medical tutor there are three definitions for medical professionals: \textit{\textbf{novice, journeyman, and master}}. Students are considered as novices and with the help of medical tutors who are in the journeyman category, they will learn how to perform palpation skills. However, the last group are domain specialists (e.g. Gastroenterologists) who are considered as masters in abdominal palpation examination skills. Hence, a light to moderate press could give them enough information about the potential abnormality in a deep palpation examination (see fourth tutor's performance in deep palpation figure \ref{fig:quartiles}) but examination with the same force magnitude might be insufficient for a medical student (novice) to diagnose the condition correctly. Figure \ref{fig:journey-master} illustrates a comparison between journeyman (third tutor) and master (fourth tutor) performances in our study.

\subsubsection{Variations across Genders and Body Types}
\begin{figure}[htbp]
\centering
\subfloat[Superficial Palpation]{\includegraphics[width=0.5\textwidth]{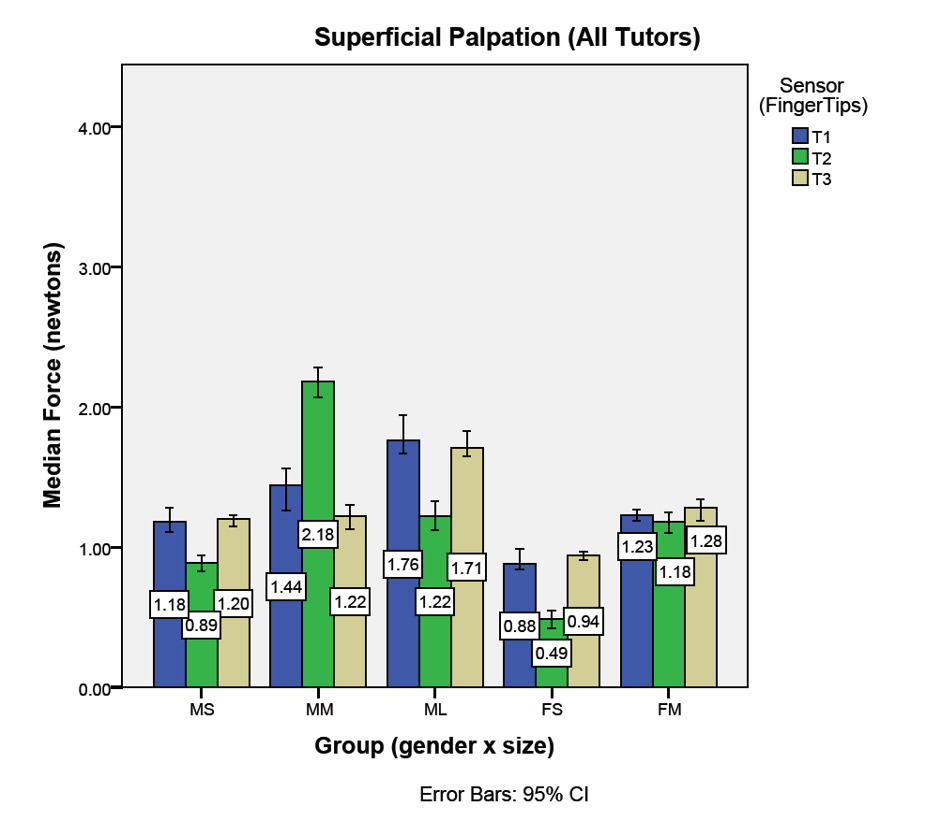}}
\subfloat[Deep Palpation]{\includegraphics[width=0.5\textwidth]{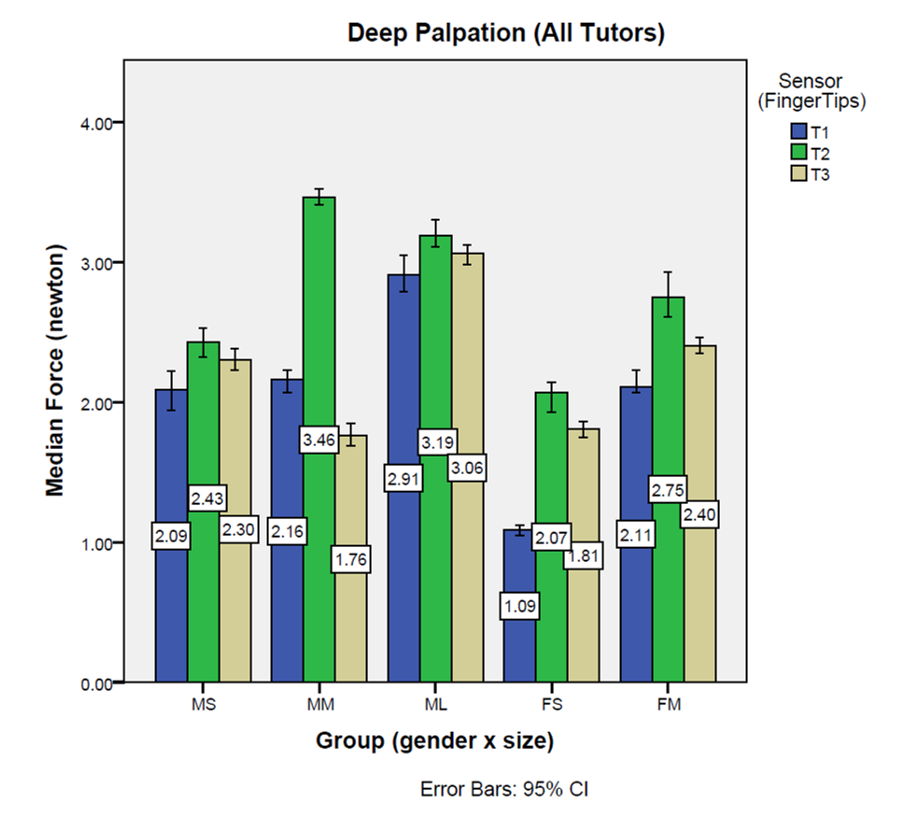}}
\caption{Finger-tip sensors' median force (in Newtons) for different body types across genders when all three sensors were engaged}
\label{fig:sup-dep-gs}
\end{figure}

Slightly higher forces were recorded on the finger-tip sensors ($T_1,T_2,T_3$) for male actor patients during superficial and deep palpation examinations by the lead medical tutor. This could be explained by the anatomical variations between male and female. The required force to detect organs and abnormalities has a positive correlation with abdominal obesity because of a thicker layer of fat around the abdomen. This means harder presses are needed on a larger anatomy in both superficial and deep palpation tasks. Figure \ref{fig:sup-dep-gs} shows the finger-tip sensors' mean force (in Newtons) in superficial and deep palpation examinations for different body types across genders.

\begin{figure}[htbp]
\centering
\mbox{\subfloat[1st Tutor]{\includegraphics[width=0.5\textwidth]{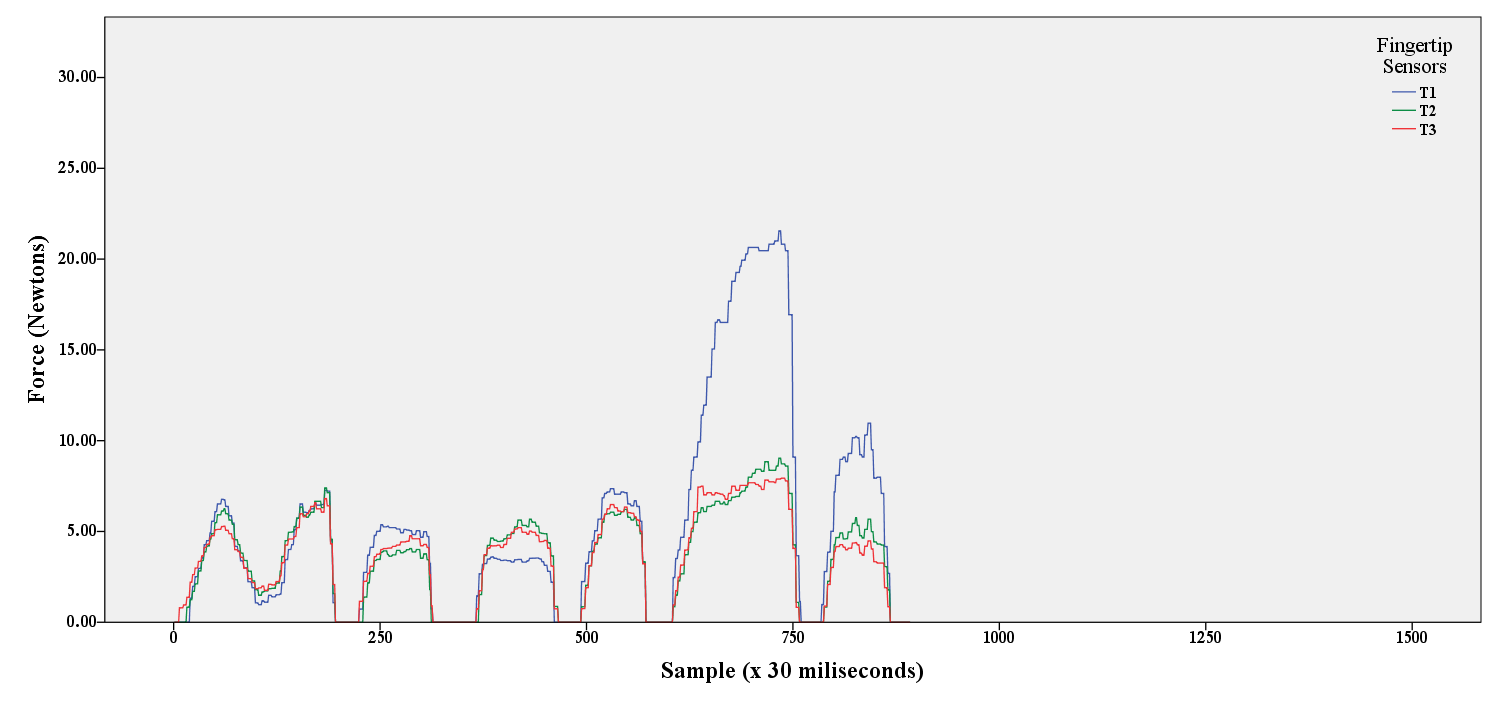}}
\subfloat[2nd Tutor]{\includegraphics[width=0.5\textwidth]{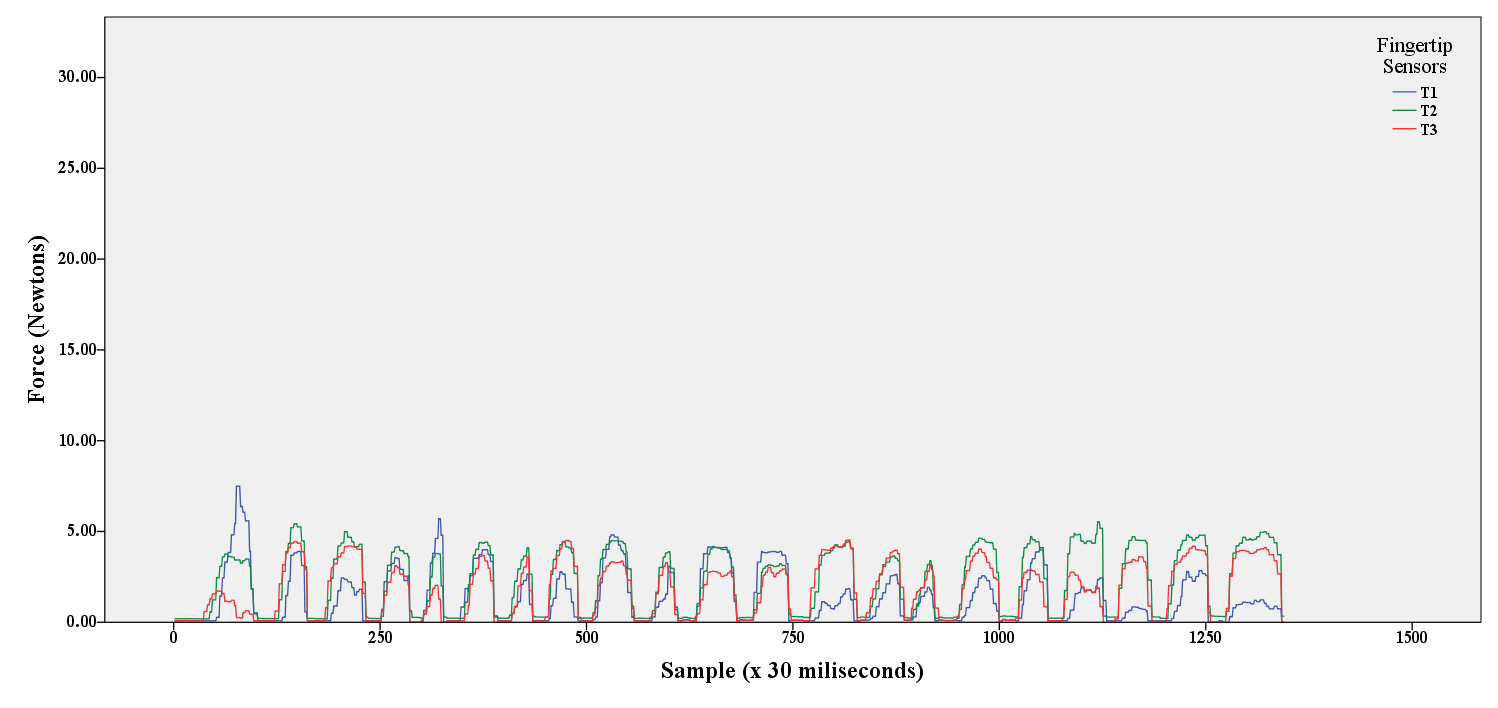}}}
\mbox{\subfloat[3rd Tutor]{\includegraphics[width=0.5\textwidth]{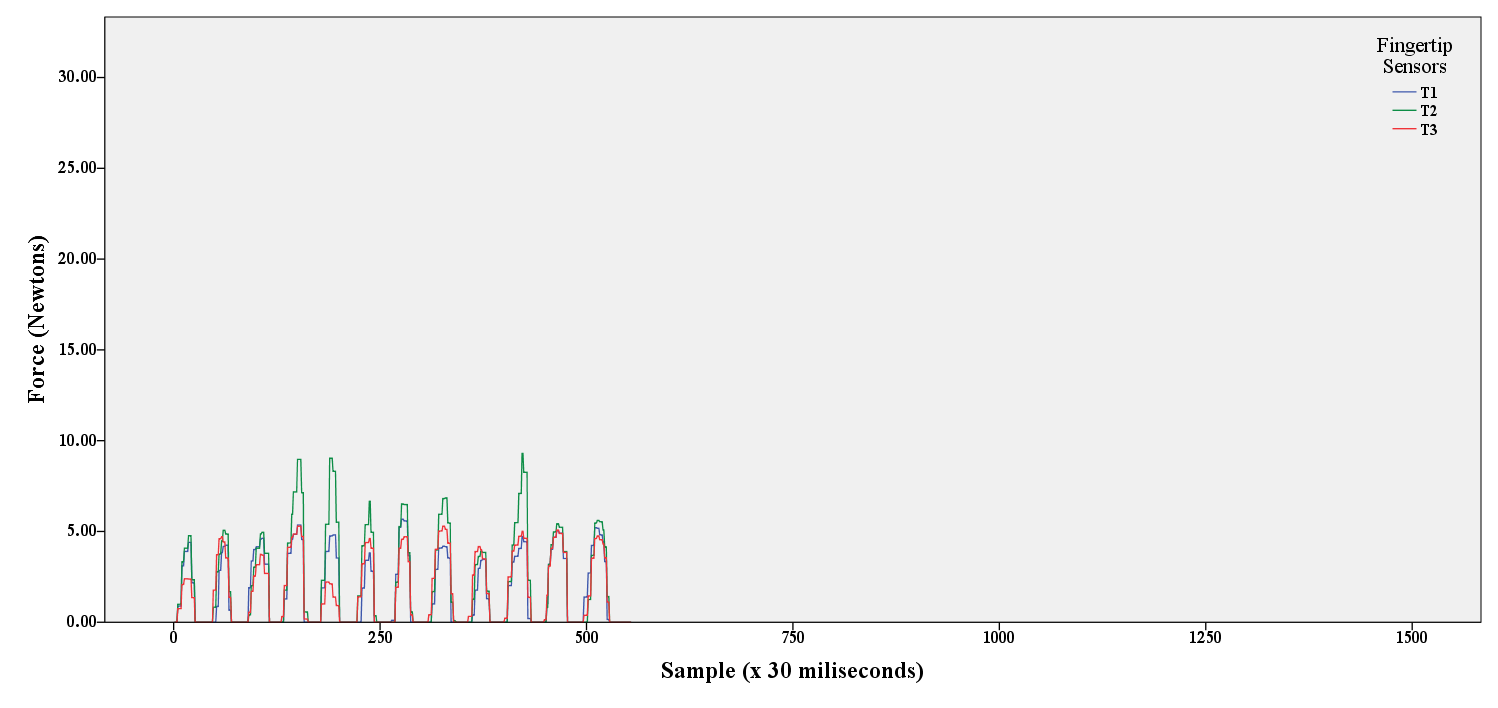}}
\subfloat[4th Tutor]{\includegraphics[width=0.5\textwidth]{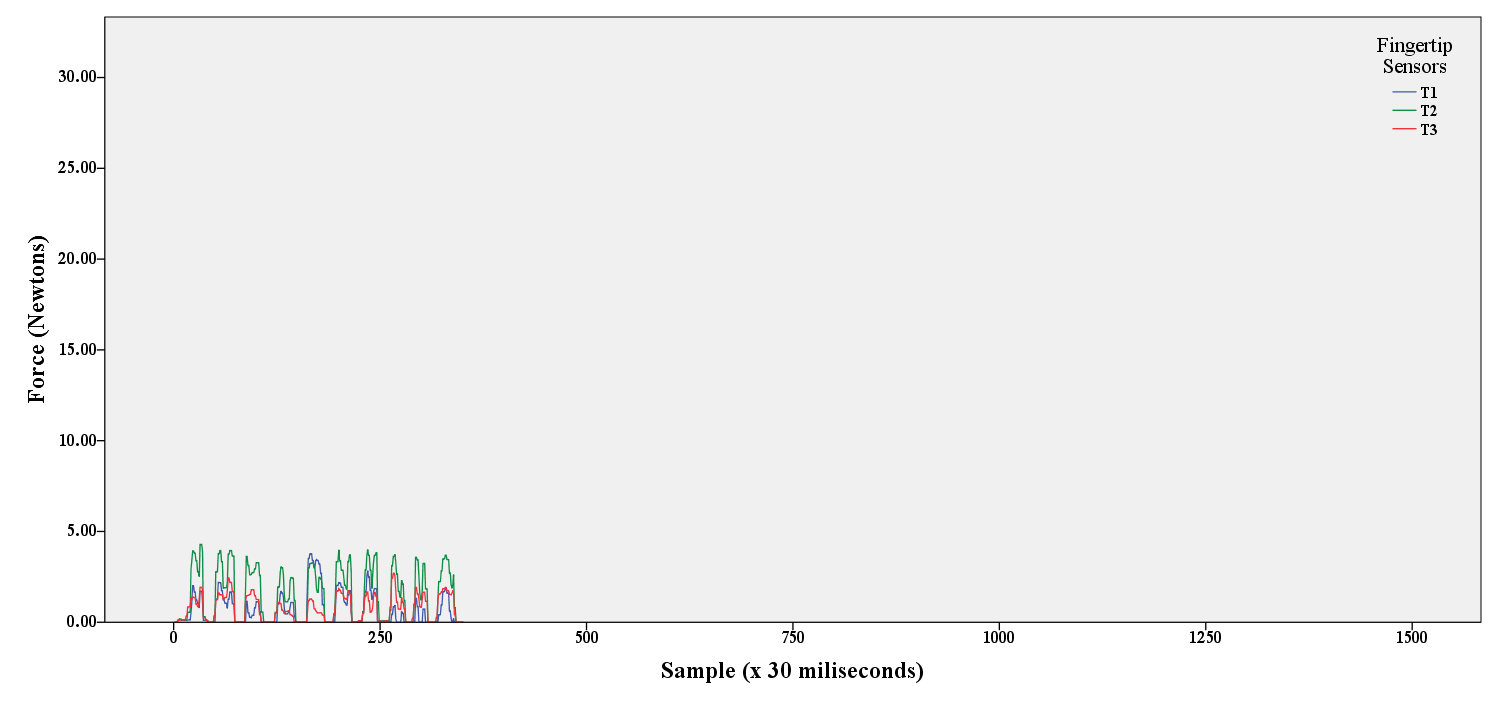}}}
\caption{Variation in deep palpation performances amongst the four medical tutors}
\label{fig:all-tutors-deep}
\end{figure}

Also, the tutors' performance in deep palpation examination (finger-tip sensors ($T_1,T_2,T_3$), large male category) were plotted to illustrate the variations between palpation techniques among tutors in current practice. According to the generated force-time plots which are presented in figure \ref{fig:all-tutors-deep}, the first tutor had harder presses for longer durations compared to all other experts with fewer regions to examine (six press-release actions). The second tutor had the longest examination (twenty one press-release actions) with moderately equal amount of force in each. The press-release durations and subsequently the total examination time were arguably decreased in the last two performances. Finally, the last tutor (the Gastroenterologist) had the lowest force readings compared to the others, close to his superficial levels.

How to resolve the variations between professionals is out of scope of this paper but it will be investigated in the future to develop a highly accurate palpation model. The following mean force levels were identified for the index finger-tip sensor; superficial palpation ($\mu= 1.25$ N), deep palpation ($\mu=2.37$ N), and lowest maximum force for deep palpation as the highest safe threshold for a small-female ($\mu=1.65$ N).


%
\section{Evaluation Study}
\label{sec:evaluation}
The final stage of this work involves the evaluation of the impact of the proposed
 training and assessment technique in real-world practice.
 This section presents a pilot study that demonstrates the potential of this work within a formal training session.

\subsection{Design}
A between-subject design was chosen for this experiment. A systematic random sampling method \citep{schutt2011investigating} was used to assign participants to one of the following three groups: A \textit{\textbf{Conventionally Trained (CT)}} control group ($n=8$) that performed palpation examination in a conventional fashion, a \textit{\textbf{Semi-visually Trained (SVT)}} group  ($n=8$) in which augmented feedback was delivered only during the training phase, and finally a \textit{\textbf{Visually Trained (VT)}} group ($n=7$) where augmented feedback was provided in both training and actual test phases.

The motivation of this experiment was to explore the effect of additional sensory information (e.g. visual feedback on applied forces) via different learning styles on cognitive and motor-control abilities to highlight any potential improvements in learning and performance. This effect was evaluated by comparing the students' competency in abdominal palpation examination with the best practice model,  proposed in the previous section, as a trusted reference model.

Our main hypothesis in this experiment ($H_1$) was that additional sensory feedback on applied forces by the medical students' hands would improve their competency in abdominal palpation examination. The null hypothesis ($H_0$) was set to ``no difference between groups in their overall performance scores’’.

The independent variable is the learning style which varies between groups. The dependent variable is based on an assessment score given to the students depending on their performance. Details of how the assessment was computed are given in the following section. Table \ref{table:doe2} shows an overview of the experimental design and different methods of training and test phases between groups.

\begin{table}[htbp]
\caption{Evaluation Study - Experimental Design}
\label{table:doe2}
\centering
\def\arraystretch{1.25}
\begin{tabular}{l|l|l|l|}
\cline{2-4}

\multicolumn{1}{c|}{} &
\textbf{Control} &
\textbf{Semi-visually Trained} &
\textbf{Visually Trained}
\\ \hline

\multicolumn{1}{|c|}{\textbf{Familiarisation}} &
Visual Feedback &
Visual Feedback &
Visual Feedback
\\ \hline

\multicolumn{1}{|c|}{\textbf{Training}} &
No Feedback &
Visual Feedback &
Visual Feedback
\\ \hline

\multicolumn{1}{|c|}{\textbf{Test}} &
No Feedback &
No Feedback &
Visual Feedback
\\ \hline

\end{tabular}
\end{table}

In addition, to minimise the influence of additional factors on students' abdominal palpation competency, e.g. previous experience in target palpations, the following control decisions were made prior to the experiment:

The PTS interface was used throughout the quantification process (\textit{Same Quantification Tool}). Also, all groups had equal chance prior to their data capturing sessions to acquaint themselves with the interface in the familiarisation phase. Homogeneity in the target population was a key to recruit participants (\textit{Same Level of Knowledge and Experience}). Thus, first year medical students were chosen to ensure participants are, on average, at the same level of knowledge and experience in clinical skills. Moreover, as mentioned in the previous section, the actor patient's anatomical variations can seriously affect the force requirements for a best practice model (\textit{Same Actor Patient}). Thus, the same actor patient was involved in all palpation sessions.

Finally, a feedback form was prepared in two parts (quantitative and qualitative). The first section of the feedback form was adapted from the System Usability Scale (SUS) concept with close-ended rating scales to let the students rate usability of the new learning style. Open-ended questions were presented in the last part of the form to collect students' feedback about their experience in a self-reporting fashion. The usability feedback form was given to the participants in the SVT and the VT groups ($n=15$) since they have had a chance to try the technology-aided method as a learning tool.

\subsection{Competency Assessment}
Students' competence in performing target abdominal palpation examination tasks is defined as their performance. Medical tutors currently assess students' performance by skills observation and question/answer techniques as part of a conventional assessment method. As mentioned previously, an assessment criteria was designed and proposed with the help of a lead medical tutor.

\subsubsection{Computer-based Assessment Method}
A second version of the report was generated by \textbf{\textit{Examogram}} using a computer-based assessment approach to visualise the analysis results for different criteria in the assessment framework. Apart from the last criteria in which only two palpation tasks were assessed (Superficial and Deep), the overall score is computed by averaging scores for all three tasks. In this method ten points in each category was awarded if the target assessment criteria was met otherwise the absolute error was computed to reduce the final score.

\begin{figure}[htbp]
\begin{minipage}[c]{.5\textwidth}
\centering

   \subfloat[Participant 22, CT group]{
   \label{fig:ctrl-2nd-rep}
   \includegraphics[width=.95\textwidth]{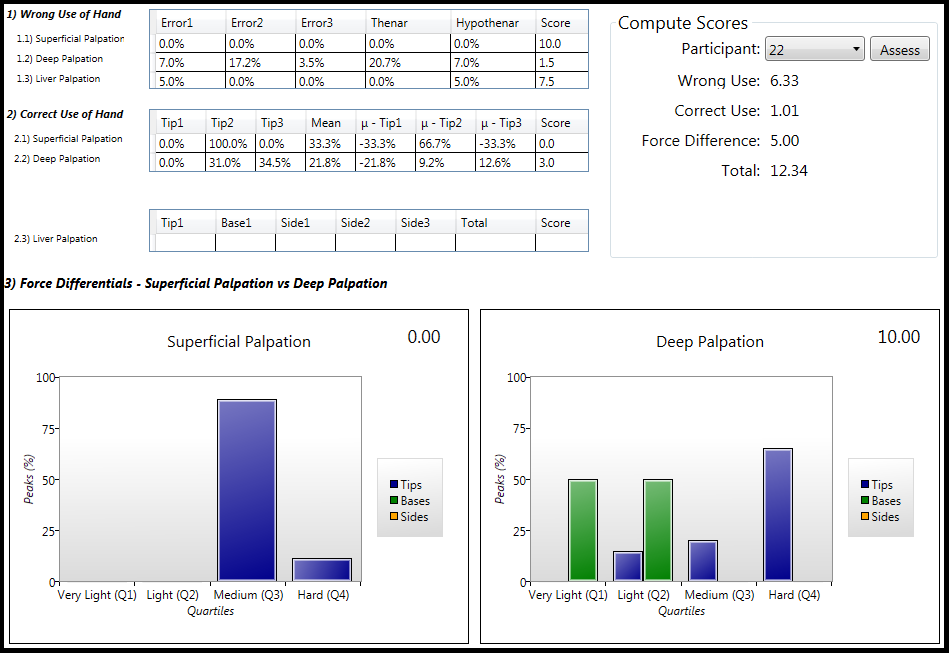}}\par

   \subfloat[Participant 11, SVT group]{
   \label{fig:semi-2nd-rep}
   \includegraphics[width=.95\textwidth]{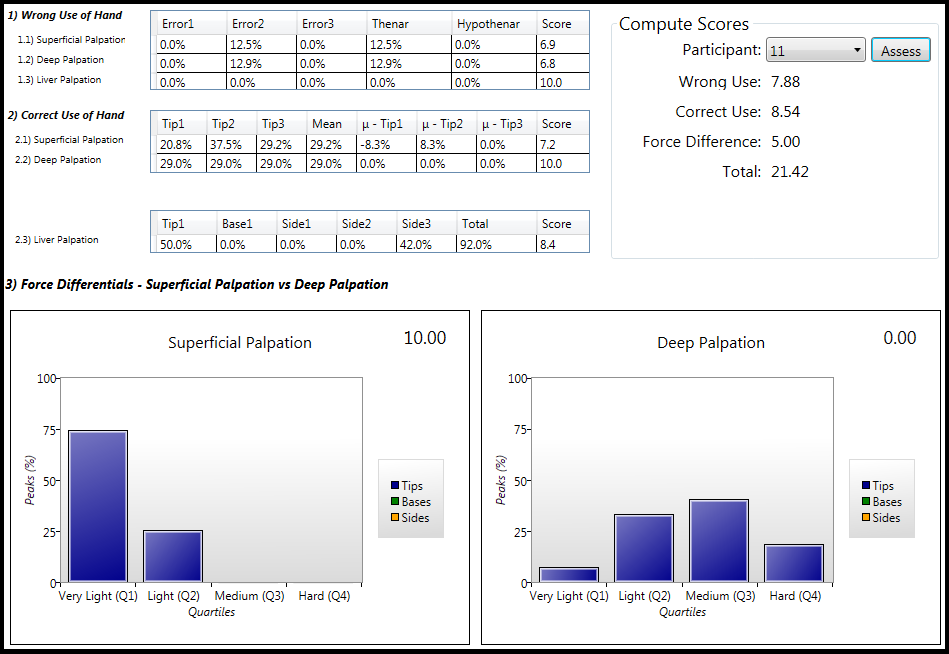}}

\end{minipage}%
\begin{minipage}[c]{.5\textwidth}
\centering

   \subfloat[Participant 9, VT group)]{
   \label{fig:visual-2nd-rep}
   \includegraphics[width=.95\textwidth]{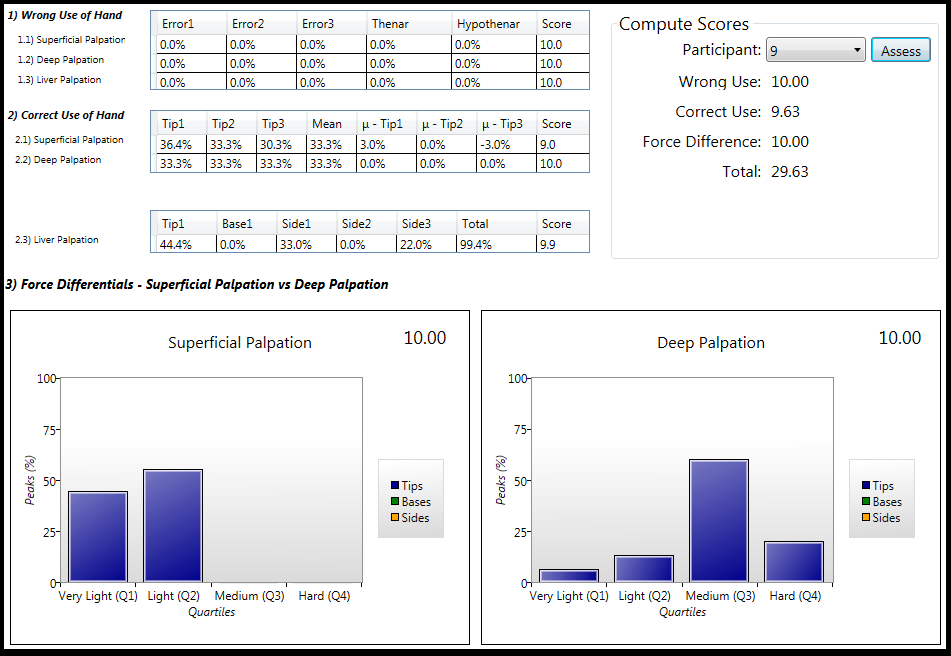}}

\end{minipage}%

 \caption{Computer-based assessment reports for control (a), semi-visually (b) and visually trained (c) groups}
 \label{fig:comp-rep}

\end{figure}

Three randomly selected competency assessment reports that were generated in computer-based assessment phase are presented in figure \ref{fig:comp-rep}.

\subsubsection{Human-based Assessment Method}
\begin{figure*}[t]
    \centering
    \includegraphics[width=\textwidth]{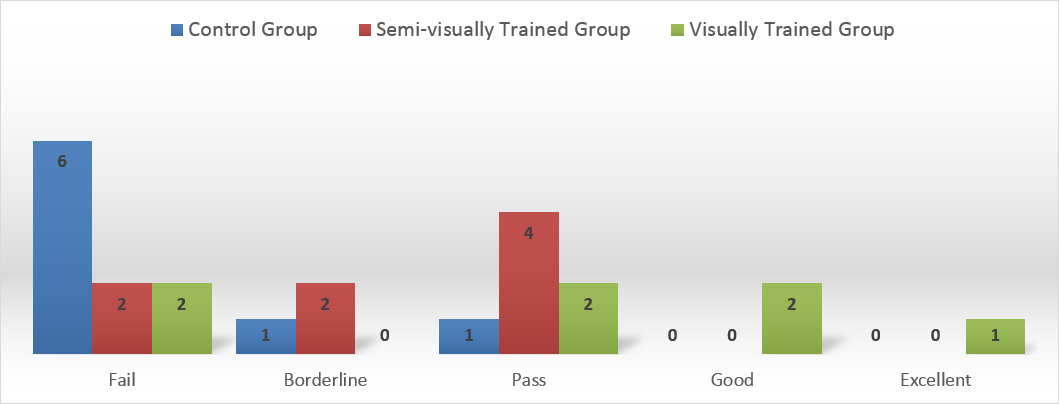}
    \caption{Tutor's combined decision in human-based assessment session}
    \label{fig:osce-ranks}
\end{figure*}

The computer-based assessment second reports were given to two medical tutors for a human-based assessment session. Tutors were not aware of students' names or their group and computer generated scores were also removed to avoid any bias. All reports were individually assessed by the two medical tutors and a final score out of thirty was given to each participant. The final score then converted to a representative categorical value (\textit{Fail, Borderline, Pass, Good, Excellent}) according to the Objective Structured Clinical Examination (OSCE) global rating template \citep{Mrcs2015}. Finally, the two medical tutors made a combined decision for each participant by reviewing scores together. A one-to-one feedback was provided to those participants who failed or their score was borderline to highlight their strengths and weaknesses. Figure \ref{fig:osce-ranks} shows the final outcomes for three groups.

\subsection{Materials}
The primary materials used correspond to the technologies discussed in section \ref{sec:interface}. Although all three previously defined palpation metrics for reconstruction of a hands-on practice (\textit{Position, Orientation, Pressure}) were collected from PTS interface, this study focused only on the force related information.

\begin{figure}[htbp]
\centering
\includegraphics[width=0.7\textwidth]{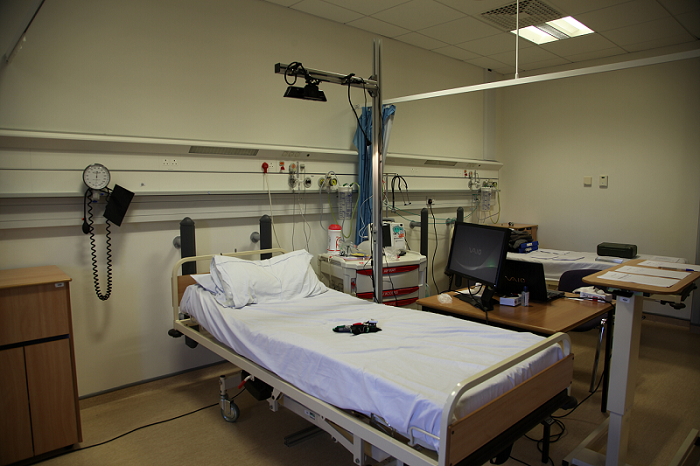}
\caption{Evaluation Study - Environmental Settings (University Hospital Coventry and Warwickshire) }
\label{fig:expii-env}
\end{figure}

The PTS-DAQ was used on a laptop to capture raw digital outputs of the twelve force sensors from the ParsGlove wearable sensory input. Force-related metrics were illustrated on a secondary panel for technology-aided groups (see figure \ref{fig:expii-env}).

The laptop monitor was positioned in front of the investigator to let him monitor the experiment's progress on PTS-DAQ primary application panel and it was obscured from participants during the whole experiment (see figure \ref{fig:pts-daq}). Examogram was used to run the preliminary queries on the captured data to present the information on the instant electronic reports for further evaluation by the medical tutors (see figure \ref{fig:examogram}).

A TFT display was used in extended mode to provide visual feedback for participants. This monitor was positioned in front of each participant to visualising the glove sensors' force magnitude and its location on the hand (see figure \ref{fig:pts-daq}). \rvm{The students are always advised to look at the patient’s face for any signs of discomfort. The investigator had limited permissions to change the room settings to position the TFT display next to the patient's face. However, PTS will deliver feedback via Augmented Reality (AR) displays in future to avoid this problem.}    

A Kinect v1.0 sensor was attached to an off-the-shelf mounting rack to provide 90 degrees base rotation in order to capture the patient's abdomen from the top. Figure \ref{fig:expii-env} shows the experiment settings.

A hospital bed with the ability to electronically adjust the height was used in the training room. To meet hygiene and safety requirements in this study, an ultra thin powder coated polyvinyl glove was provided to each participant before asking them to wear ParsGlove to avoid skin irritations and electrostatic shocks. Moreover, a medically approved sanitising hand gel was used on the ParsGlove surface before each examination similar to the real-world practice to avoid skin irritations or infections.

\subsection{Participants}
Twenty three participants took part in this experiment with twelve females and two left-handed subjects. Except for the \textit{\textbf{Visually Trained}} group ($n=7$) each group had eight participants. They all had normal or corrected to normal vision. All participants were chosen from first year medical students in WMS  to minimise the influence of having prior knowledge and experience on students' competency results. The investigator's colleague took part in this experiment as the actor patient.

\subsection{Procedures}
Prior to each session a brief description of the experiment was provided in four Power point slides. In each trial only three people were in the experiment room: the investigator, the actor patient, and the medical student. Each participant was greeted by the investigator. A signed consent form was taken from each participant and an email was voluntarily provided for electronic reports.

\begin{figure}[htbp]
    \centering

    \subfloat[CT group]{\includegraphics[width=0.5\textwidth]{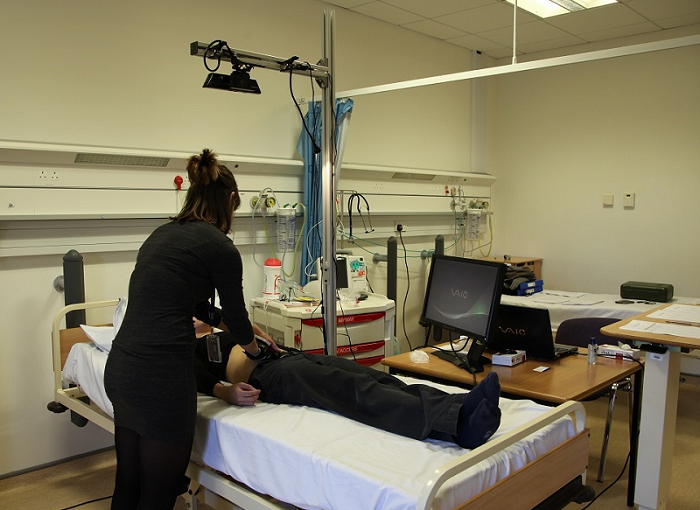}}
    \subfloat[SVT group]{\includegraphics[width=0.5\textwidth]{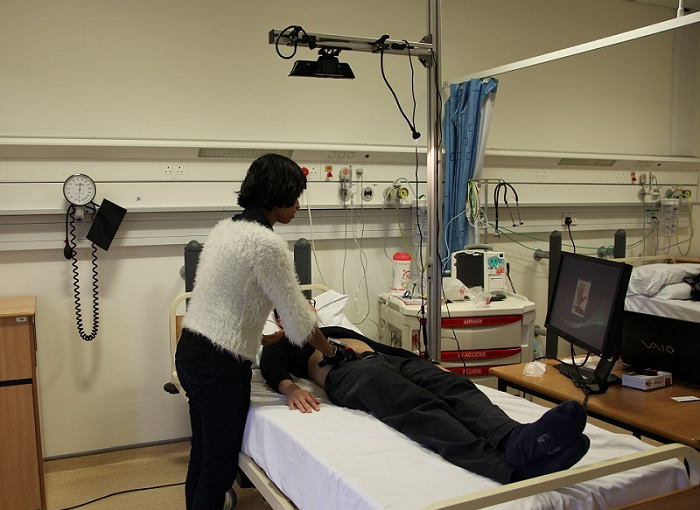}}

    \caption{Experiment room and two different learning methods to teach abdominal palpation examination.}
    \label{fig:exp}
\end{figure}

A training phase and an actual test phase were planned for each target abdominal palpation examination task. Different learning styles were employed in each phase according to the participant's group. In addition, they were asked to conduct the examination session for each task similar to its real-world practice. Figure \ref{fig:exp} shows the experiment room and different learning styles.

Participants began by introducing themselves to the actor patient followed by a brief explanation on the examination process. The data collection process was initiated and terminated for each abdominal palpation task by the students' verbal signal. Finally, raw data samples were labelled and stored on the investigator's laptop.
The actor patient had been briefed by the investigator about the experimental procedure at the start.

\subsection{Results}
A mixed mode assessment approach is used in this study to highlight the impact of the new learning method on student's palpation performance and overall experience via quantitative and qualitative research methods.

\subsubsection{Performance Assessment}
A non-parametric \textit{Kruskal-Wallis} test was used on categorical results (OSCE ratings) as an equivalent counterpart for \textit{one-way ANOVA} to ensure the results are not affected by the small size of samples. Also, equal variances were reported amongst groups, $F(2,20) = 2.240$, $p < .05$ by a Levene's test on assessment ranks. Students' abdominal palpation performances were reported to be significantly affected by the presence of visual feedback on applied forces $H(2) = 6.033$, $p < .05$. Thus, the null hypothesis ($H_0$) is rejected.


A post-hoc pairwise comparison test using \textit{Bonferroni correction} revealed that the mean score for training and test phases in the \textit{VT} group (M = 2.00, SD = 1.53) significantly differed from the \textit{CT} group(M = .38, SD = .74). However, providing the new learning method only in the training session in \textit{SVT} group (M = 1.00, SD = .76) did not show significant difference from the other two conditions (\textit{CT} and \textit{VT} groups).

\begin{figure}[htbp]
\centering
\includegraphics[width=0.6\textwidth]{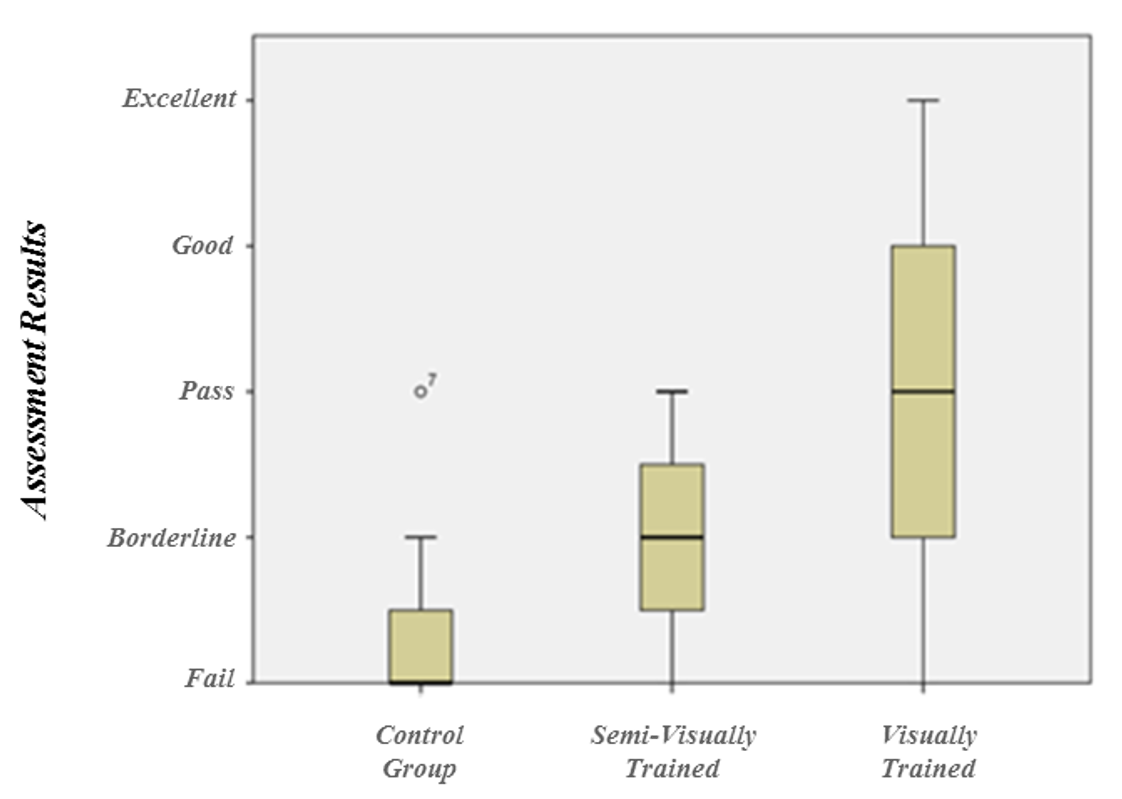}
\caption{Gradual improvement in the medical students' palpation performance (between subjects)}
\label{fig:eval}
\end{figure}

Also, positive trend in students' performance improvement is reported by \textit{Jonckheere-Terpstra} test results when visual feedback on applied forces was used in palpation sessions. The median rank increased in groups, $J=132$, $z=2.62$, $r=0.55$. Figure \ref{fig:eval} shows this trend.

A significant positive correlation is reported between computer-generated scores and human-generated scores using \textit{Pearson Product-Moment Correlation}, $r=.637$, $p$ (one-tailed) $< .05$. This indicates the potential of the computer-based assessment technique to assist current assessment process in medical education.

Although, both scores were generated according to the assessment criteria, the important role of tutors in assessment process is clearly evident in the final results. For instance our last participant in \textit{VT} group has achieved a remarkable score of $23.15$ in computer-based assessment method and $19.5$ in human-based assessment method out of thirty but since the correct use of hand in palpation of liver edge indicates no reading from target sensors at all a final decision of \textit{Fail} is given in a combined decision by tutors.

\subsubsection{Students feedback on Method's Usability}
\begin{figure*}[t]
\centering
\includegraphics[width=0.9\textwidth]{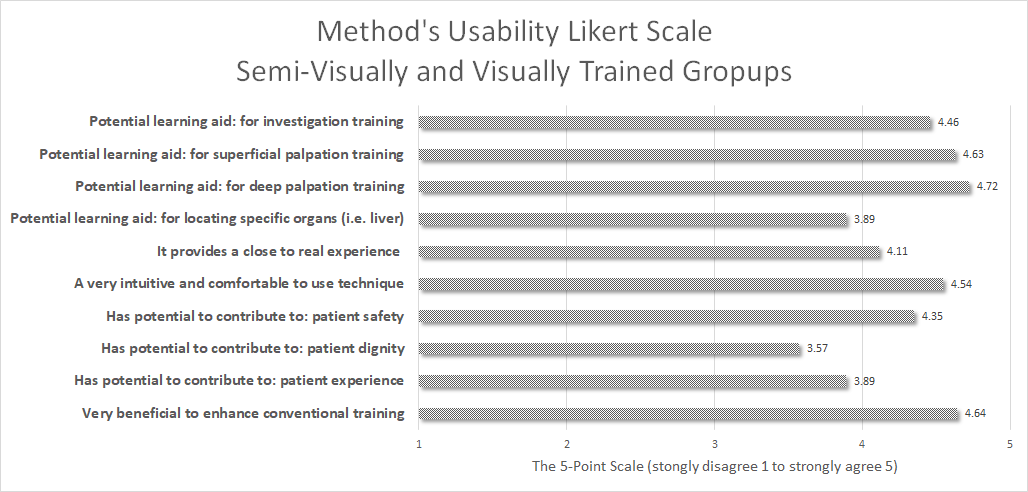}
\caption{Students' feedback on usability of the proposed technique}
\label{fig:eval-likert}
\end{figure*}

Eleven feedback forms out of fifteen were received from the students comprised of seven females with one left handed subject. The results of subjective assessment by the medical students in the SVT and the \textit{VT} groups are presented here:

\paragraph*{Likert Scale}
In first part of the form, students were asked to choose the corresponding number for the most relevant answer. Figure \ref{fig:eval-likert} shows the means for the five point scale rating system (strongly disagree 1 to strongly agree 5) which was used in students' questionnaire.

The vast majority ($81\%$) of the participants in this survey ($n=11$) from the technology-aided groups (\textit{VT} and \textit{SVT}) who have returned their feedback forms, agreed that this method has a potential in leaning abdominal palpation examination skills. The potential usability of this method is also surveyed in a task-specific level in which majority of students agreed its potential ( ($89\%$) for Superficial and ($94\%$) for deep palpation). One good reason for a better score in deep palpation could be the necessity of a clear definition for a deep press to avoid any hesitation. However, a lower score of ($46\%$) is achieved for locating organs in the third target abdominal palpation task (liver palpation) which could be explained by the wearable interface. Fifty eight percent of the students have described their experience as close to real. Moreover, the measurement interface in this learning method was described as very comfortable and very light-weight ($85\%$). It was also stated as a potential tool to contribute to patient safety ($70\%$) but less likely to contribute to patient dignity ($35\%$) and to patient experience ($45\%$). Finally, a very promising score ($90\%$) was given to this method as a beneficial learning approach to enhance conventional abdominal palpation examination practice in current medical education. Since only one left-handed person was participated in this survey, it neither agreed nor disagreed that left/right handedness may affect learning these skills.

\paragraph*{Questionnaire}
The students gave answers to five open-ended questions \rvm{(learning outcomes, strengths, weaknesses and other comments)} in the second part of the feedback form in a self-reporting fashion. Written answers to the questions were summarised in the following two categories:

\begin{itemize}

\item \textbf{Learning Outcomes:} The answers to the first question highlight better understanding of force differentiation for each abdominal palpation task. One student explicitly discussed about potential hesitation in palpating hard and deep enough due to lack of a clear definition for deep palpation. In addition to that, a better understanding of correct and effective use of the palpating hand per each abdominal palpation task was indicated among the answers.

\item \textbf{Strengths and Weaknesses:} Properties such as enhancing the palpation accuracy, contribution to patient's safety, instant feedback on applied forces, navigation based on tutor's best practice, ease of use and high sensitivity to application of force by the palpating hand were discussed as key strengths of this method. Potential weaknesses were also identified among the answers. \rvm{Majority of their comments were focused on design and room settings. In particular, the following concerns raised among the answers; Paying attention to the patient's face during familiarisation stage, glove size for small hands and left handedness, exposure of the electronics to patient's skin.} 

\end{itemize}


\rvm{In addition, the actor patient's \textit{(graduated medic, male, 36, BMI of ~24.5)} feedback was collected to highlight his experience. His participation added more value to the observation process. The medical students were monitored for variation of techniques and their performance in conducting examination process. In this section brief review on actor patient's experience is presented.}

\begin{itemize}

\item {\rvm{Feedback on the measurement interface: On examination (superficial, deep, and liver palpation), the sensation of the glove on the abdominal skin is stated as comfortable with no hard or sharp edges or excessive friction when used by a competent examiner. In addition, the motion capturing part of interface ({Kinect\textregistered} v1.0 and aluminum frame) was described as non-intrusive with ability to comfortably lie flat beneath it.}}

\item {\rvm{Feedback on the examination sessions: Despite having just attended a second clinical examination skills tutorial, notable variation in the performance and a few competence issues were observed during participants' examinations. One or two participants were described as very confident on examination and a lot more skilled than the others. A few needed reminders on examination technique with two subject who applied a completely wrong technique in the palpation of liver edge. This may be due to a lack of confidence, enough practice, and/or previous related experience in palpation examination. Otherwise examination routine was descried as gentle with the participants checking if the examinee was uncomfortable (i.e. good bedside manner).}}

\end{itemize}


\subsection{Discussion}
The results show a significant improvement in learning abdominal palpation examination skills which confirms the benefits a technology-aided approach can have to enhance conventional teaching methods. Moreover, potential benefits of such a method were stated in the obtained usability feedback in the qualitative evaluation. Students in both groups explicitly show their motivation to use this technique as part of their education. A few of these comments are outlined here:

\begin{quote}

`Very good tool which I think has the potential to improve abdominal palpation skills a lot. Excellent idea. I hope it is used more widely in the future.' (\textit{female,VT group})

\rvm{
`That my deep palpation wasn’t as deep as some of the others' (\textit{male,SVT group})
}

\rvm{
`Not to be afraid to palpate more deeply + risk discomfort for the patient'(\textit{female,VT group})
}

\rvm{
`I learned the parts of my hand that I wasn't using as effectively as other parts'(\textit{male,SVT group})
}

\rvm{
`Allow you to compare what you are doing with that of an expert and adjust your technique accordingly.' (\textit{female,VT group})
}

\end{quote}


%

\section{Conclusions and Future Work}
Clinical palpation skills are a cheap yet very effective method of diagnosis. Thus, it is crucial for medical students to master these core skills and to ensure that they are retained throughout their career. This is very difficult to achieve without frequent and realistic rehearsals not only during academic training hours but also in the medical students' self-study time outside of the classroom. This paper has presented a multimodal technology-aided technique to enhance conventional training and its assessment in medical education. An extensive literature review on mechanical, biological, psychological, and technological aspects of human motor learning particularly in practical hands-on interactions was followed by a comprehensive ergonomics study to understand the user and task requirements for accuracy and reliability of the method. Also, active involvement of medical experts (masters) and students (novices) in a UCD fashion was a key feature of this study to accurately design the research methodology and to evaluate its impact through real-world practice.

Palpation metrics were identified by medical tutors and a proof of concept model was developed by averaging captured palpation data from a team of four medical tutors in order to minimise the variations in captured data. Moreover, an assessment criteria was defined with the help of the lead tutor based on the proof of concept model.

Abstract visualisation on location and magnitude of exerted forces by the medical students' palpating hand were provided to semi-visually trained (feedback on training), and visually trained (feedback was provided in both training and test trials) groups to compare their performance with participants in the control group (without any feedback). The captured data from medical students during the actual test trial were assessed in a machine-based and a human-based approach using the previously defined criterion. Participants in the visually-trained group significantly outperformed the other groups ($H(2) = 6.033$, $p < .05$). The semi-visually trained group performed better than the control group but it was not significant ($p > .05$). The machine-based assessment results were in a positive correlation with the human-based results. Positive responses were received from the medical students who had been trained by the multimodal technology-aided method.

The authors are not claiming that this is the only way to achieve competence since competence will be achieved in a lifelong journey. This technology provides a platform to allow students to compare their methods with others and to learn from their mistakes. It will also provide an intelligent autonomous assessment method to suggest improvements. The intention of this research is to enhance learning and retention levels by providing more access to realistic scenarios via multimodal simulations in future. It is our intention to suggest various techniques for particular tasks and leave it to students/experts to decide which method(s) suits them best.
Moreover, traditional assessment during formal examination is subjective purely based on observing students and asking questions to interpret haptics sensations into verbal descriptions during OSCEs. This method proposed in this paper will assist tutors to measure the students’ accuracy in performing these skills in a quantitative fashion.

\subsection{Limitations}
There are still a number of limitations to our approach.

\textit{\textbf{Ergonomics Method:}} As noted in section \ref{sec:ergonomics} ethnography (or field study) is one of the most effective methods in ergonomics studies to determine challenges and collect information about conventional practice. However, it is a costly and time consuming method in which the investigator is physically involved/embedded in the whole training programme (in medical school) for long-term investigations. As part of the future work participation in several formal training sessions is planned in order to extend the current knowledge about palpation skills.

\textit{\textbf{Sampling and Data Collection:}} The proposed interface targets novice clinicians only at the very early stages in their journey towards full competency on clinical examination. The device provides objective evidence as to whether the novice student is achieving standards in superficial and deep palpation that approach the method adopted by the senior clinical tutors. Formal competence can only be gained by how the students assesses and diagnosis a real patient in a real clinical situation. Four senior clinical tutors were involved in order to develop a ground theory in a limited timeframe and more collaboration from experts are required to improve efficiency and reliability of the proposed approach.

\textit{\textbf{Development and Implementation:}} Measuring the human ergonomics demands advancement in sensory technologies and multimodal visualisation displays. Thus, more research is needed in this domain to guide future researchers and developers towards robust and reliable quantification techniques.

\subsection{Future Work}
Future work will look into enhancing this study by provision of a haptic-enabled visualisation technique (e.g. a virtual or augmented reality palpation simulator) in future evaluation studies to navigate the medical students' hand movements (the capability of the system to capture location and orientation) as well as providing feedback on force applications (which is proposed in this work). A new version of our wearable solution, an advanced hybrid tracking system, and an online solution with AI algorithms are currently under development. The online  database will be available to all medical experts who teach palpation skills to remotely acquire their palpation performance with the ability to compare it with other medical experts as an educational approach for seniors. This will also improve the accuracy and reliability of the ``gold standard model’’ and minimise the variation between tutors' techniques. In addition, we intend to investigate the impact of our novel training method in other palpation tasks.

Longitudinal studies are planned to evaluate the learning and retention levels in multiple transfer tests (withdrawal of the augmented feedback) with short- and long-term intervals. More broadly, the gold standard database will be used in the development of a multipoint haptic display to enhance the medical students' performance in a virtual reality environment with adequately realistic kinaesthetic and cutaneous feedback in conjunction with other modalities. Furthermore, actor patients' abdominal models from various body types and genders in our proof of concept study (which is saved as 3D point clouds) are captured by the proposed measurement technique. This information could be provided as a guideline to researchers and developers to form a virtual patient anatomy with realistic deformations of the human soft tissue in haptic rendering domain. Finally, a portable measurement technique with access to the live gold standard model could increase the frequency of rehearsal which is a key feature in palpation training and enhance the medical students' experience in their self-study time with accurate assessment and supervision available in simulation from tutors.

%

\bibliographystyle{elsarticle-harv}
\bibliography{bib/references}

\end{document}